\documentclass[prl,amsmath,amssymb,twocolumn,showpacs,superscriptaddress]{revtex4-1}
\usepackage{bm}
\usepackage{amssymb}
\usepackage{colordvi}
\usepackage{graphicx}
\usepackage{color}
\usepackage{hyperref}

\newcommand{\ov}[1]{\overline{{#1}}}
\newcommand{\be}{\begin{equation}}
\newcommand{\ee}{\end{equation}}
\newcommand{\bea}{\begin{eqnarray}}
\newcommand{\eea}{\end{eqnarray}}

\newcommand{\vep}{\varepsilon}

\newcommand{\ome}{\omega}

\def\nn{\nonumber}

\begin{document}

\title{Efficiency Statistics and Bounds for Systems with Broken Time-Reversal Symmetry}

\author{Jian-Hua Jiang}
\address{Department of Physics, Soochow University, 1 Shizi Street, 
  Suzhou 215006, China}
\address{Department of Physics, University of Toronto, 60 Saint
  George Street, Toronto, ON, M5S 1A7, Canada}
\author{Bijay Kumar Agarwalla}
\address{Department of Chemistry, University of Toronto, 80 Saint
  George Street, Toronto, ON, M5S 3H6, Canada}
\author{Dvira Segal}
\address{Department of Chemistry, University of Toronto, 80 Saint
  George Street, Toronto, ON, M5S 3H6, Canada}

\date{\today}

\begin{abstract}
Universal properties of the statistics of stochastic efficiency for 
mesoscopic time-reversal symmetry broken energy transducers are
revealed in the Gaussian approximation. We also discuss how the second
law of thermodynamics restricts the statistics of stochastic
efficiency. The tight-coupling (reversible) limit becomes unfavorable,
characterized by an infinitely broad distribution of efficiency at
{\em all times}, when time-reversal symmetry breaking leads to an
asymmetric Onsager response matrix. The underlying physics is
demonstrated through the integer quantum Hall effect and further
elaborated in a triple-quantum-dot three-terminal thermoelectric
engine.
\end{abstract}


\maketitle


{\sl Introduction}.--- Nonequilibrium phenomena of macroscopic systems
are described by their responses to external perturbations and follow
the laws of thermodynamics \cite{groot}. Statistical fluctuations of a
measurable quantity  are negligible in macroscopic systems but in
very small (``mesoscopic'') systems they play an essential
role \cite{noise}. Importantly, in such a mesoscopic regime a single measurement of
a physical quantity (e.g., electrical current) does {\em not} follow the laws of
thermodynamics
(although its average over many different measurements does).
Well known examples are the Jarzynski equality \cite{jarzynski,fluct-exp1,fluct-exp2,fluct-exp3}
and the fluctuation theorem, which states that a stochastic negative entropy production
may show up, though exponentially unlikely compared to its corresponding positive entropy production process
\cite{ft1,ft2,ft3}.
An important consequence of the fluctuation theorem is that the
``stochastic energy efficiency'' of a very small engine can be larger
than the Carnot efficiency, although the average efficiency over many
measurements
is smaller or equal to the Carnot efficiency, in accord with the second law of thermodynamics
\cite{me1,me2,me3,me4,njp,vdb,vdb1,martinez}. 
The study 
of stochastic efficiency of very small energy transducers is pertinent to
understanding of mesoscopic thermoelectric energy
conversion in normal electron \cite{n1,prl1} and Cooper pair islands
\cite{fazio}, biological photosynthesis in  an individual
reaction unit \cite{scully}, and photo-mechanical energy conversion in
 quantum optomechanical systems \cite{opto}. 

Analyzing stochastic efficiency, it was recently shown that the 
Carnot efficiency is the least likely stochastic efficiency \cite{me1},
later found to be solely the consequence of the fluctuation
theorem \cite{ft} for time-reversal symmetric (TRS) energy
transducers \cite{me2}. Breaking time-reversal symmetry can shift the
least likely efficiency away from the Carnot
efficiency \cite{me2,vdb,njp}. However, little is known about
efficiency statistics of time-reversal symmetry broken (TRB)
mesoscopic energy transducers, which is the gap we want to fill in
this Letter. Another fundamental question concerns how the second law of
thermodynamics restricts the statistics of efficiency for both TRS and
TRB energy transducers. In this Letter we address these problems for
systems operating in the linear-response regime where fluctuations can
be well described within the Gaussian approximation. One of our key
results is that in the reversible (or ``tight-coupling'') limit the
distribution of stochastic efficiencies becomes infinitely broad at {\em
 all times}, thus, the average efficiency loses its meaning even in the
infinite long time limit. This anomaly occurs only for reversible TRB
energy transducers of which the Onsager response matrix is asymmetric.
We discuss this anomaly using the example of the integer quantum Hall
effect. We further demonstrate our result with a 
triple-quantum-dot (QD) thermoelectric engine where a full-counting
statistics method confirms our Gaussian theory.

{\sl Efficiency statistics in the Gaussian approximation}.--- We
consider a generic situation in which there are two energy output
channels (``1'' and ''2''). Each of the channels has a
thermodynamic ``current'' and a conjugated driving force (i.e.,
affinity). The time-integrated currents are denoted by $J_i$ ($i=1,2$)
while the time-intensive current is defined as $I_i=J_i/t$ with $t$ 
the total time of operation. The affinities are associated with
the properties of the reservoirs
(e.g., temperatures and electrochemical potentials) 
and hence fluctuate negligibly. In contrast, the
currents may fluctuate considerably. 
A small TRB machine can be characterized in the linear-response regime by
$\ov{I}_i=M_{ij}A_j$ ($i,j=1,2$), or $\vec{\ov{I}}=\hat{M}\vec{A}$ with
$\vec{\ov{I}}=(\ov{I}_1,\ov{I}_2)$ and $\vec{A}=(A_1,A_2)$.
As well, in this regime
the statistics of the currents at long time $t$ can be described within the
Gaussian approximation by the distribution 
$P_t(\vec{I})= \frac{t\sqrt{\det((\hat{M}^{-1})_{sym})}}{4\pi}
\exp(-\frac{t}{4}\delta
\vec{I}^T\cdot\hat{M}^{-1}\cdot\delta\vec{I})$ \cite{gaspard}. Here
$\det((\hat{M}^{-1})_{sym})$ is the determinant of the 
symmetric part of the inverse of the Onsager response matrix $\hat{M}$
and the superscript ``$T$'' denotes transpose. While averaged
quantities are represented with a bar over the symbols throughout this
paper, $\delta \vec{I}=\vec{I}-\vec{\ov{I}}$ represents fluctuations
of the currents. From the probability distribution of stochastic
currents we calculate the distribution of efficiency
$P_t(\eta)$ \cite{me3}. We then obtain the large deviation function
(LDF) of the stochastic efficiency ${\cal G}(\eta)\equiv - \lim_{t\to
  \infty} t^{-1} \ln[P_t(\eta)] $. 
%
The scaled LDF is derived in the Supplementary Material. It is given by 
\begin{align}
  \hspace{-0.265cm}J(\eta) &\equiv \frac{{\cal G}(\eta)}{\ov{S}_{tot}} \nn\\
  &= \frac { J(\eta_C) \left ( \eta + \alpha^2 +  \alpha q r + \alpha q
      \eta \right)^2} {(1 + \alpha^2 + \alpha q r + \alpha q)\left ( 
      \eta^2 + \alpha^2 +  \alpha q\eta +  \alpha q r
      \eta  \right)} ,\label{eq:jeta}
\end{align}
where $\ov{{S}}_{tot}=\sum_i\ov{I}_iA_i$ is the average total entropy
production rate and 
\be
J(\eta_C)\equiv \frac{4-q^2(1+r)^2}{16(1-q^2r)} \label{jetaC}
\ee
is the {\em scaled LDF at Carnot efficiency}. 
Here,
\be
q \equiv \frac{M_{21}}{\sqrt{M_{22}M_{11}}},\quad r \equiv \frac{M_{12}}{M_{21}},
\quad \alpha \equiv \frac{A_1\sqrt{M_{11}}}{A_2\sqrt{M_{22}}},
\ee
are dimensionless
parameters that characterize the responses of the system and the applied affinities.
We term $q$ the degree of coupling \cite{caplan}, $r$  the
TRB parameter \cite{casati}, and $\alpha$  the affinity
parameter \cite{jiang}. In addition, the efficiency is defined as 
$\eta=-I_1A_1/(I_2A_2)$ \cite{caplan,me2,jiang,gaspard}. For thermal engines,
$\eta=\tilde{\eta}/\tilde{\eta}_C$, with the standard definition of energy efficiency
$\tilde{\eta}=W/Q$ 
and $\tilde{\eta}_C$ the original Carnot efficiency. In our scheme,
efficiency is scaled so that the Carnot (reversible) efficiency corresponds to
$\eta_C\equiv 1$.

\begin{figure}[]
\begin{center}
\hspace{-0.25cm}\includegraphics[height=3.7cm]{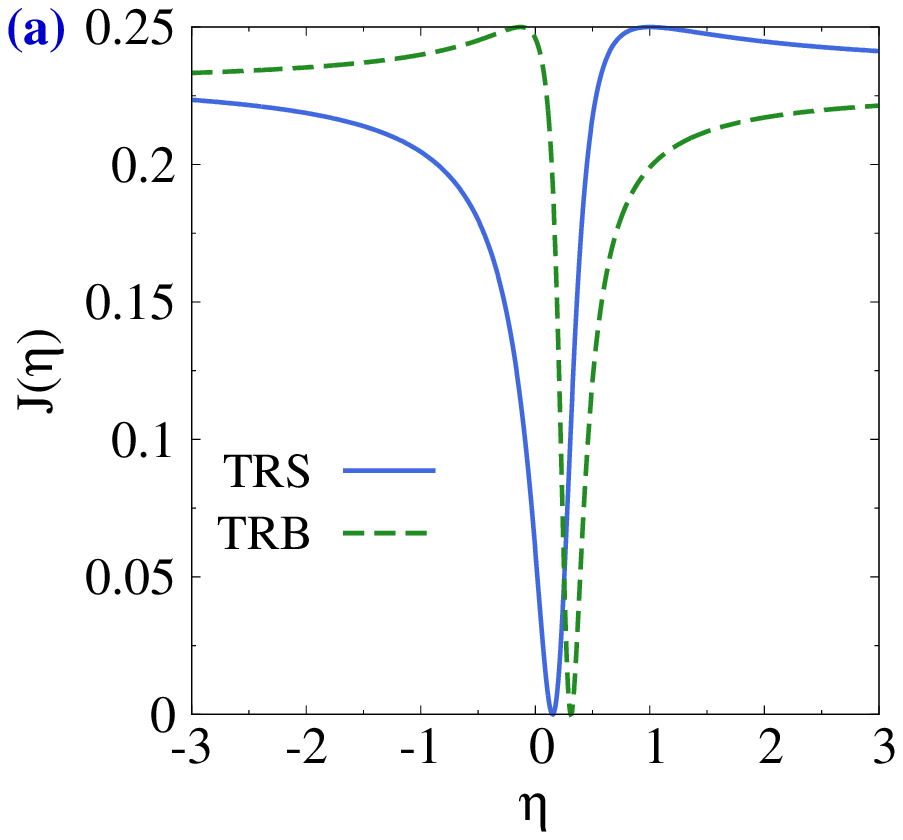}\includegraphics[height=3.7cm]{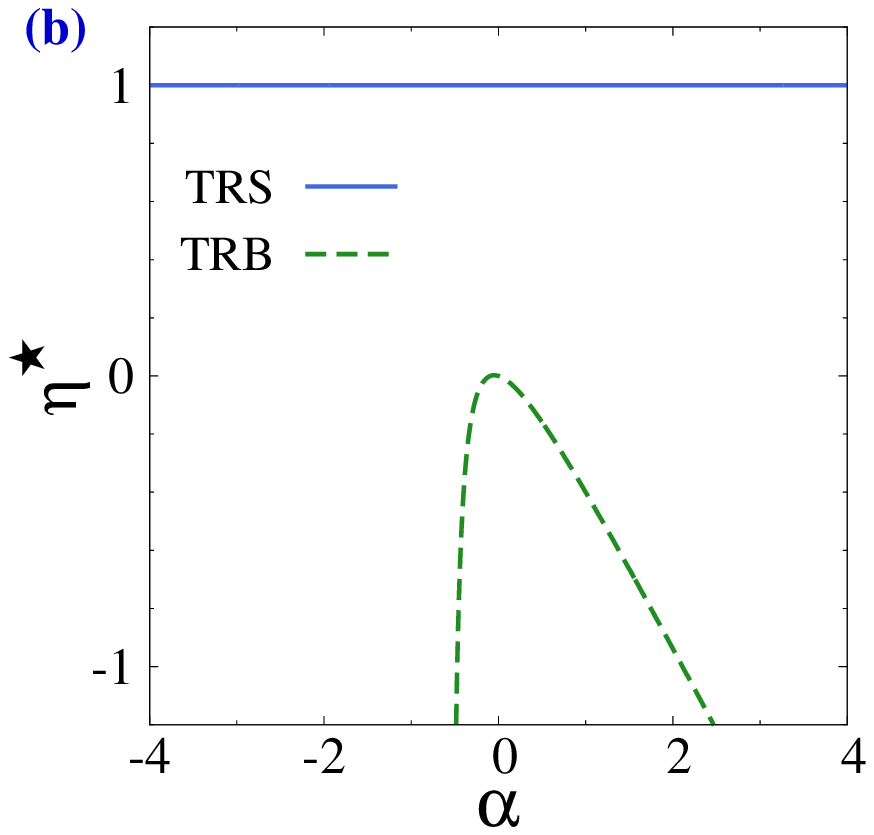}
\includegraphics[height=3.7cm]{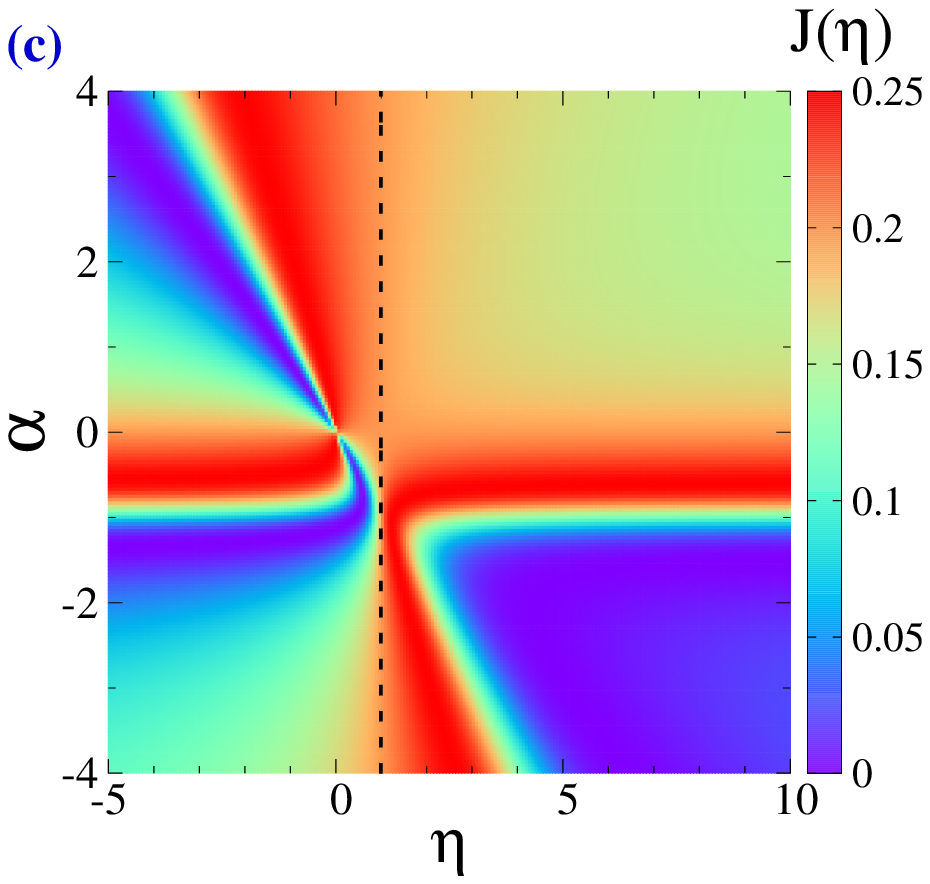}\includegraphics[height=3.7cm]{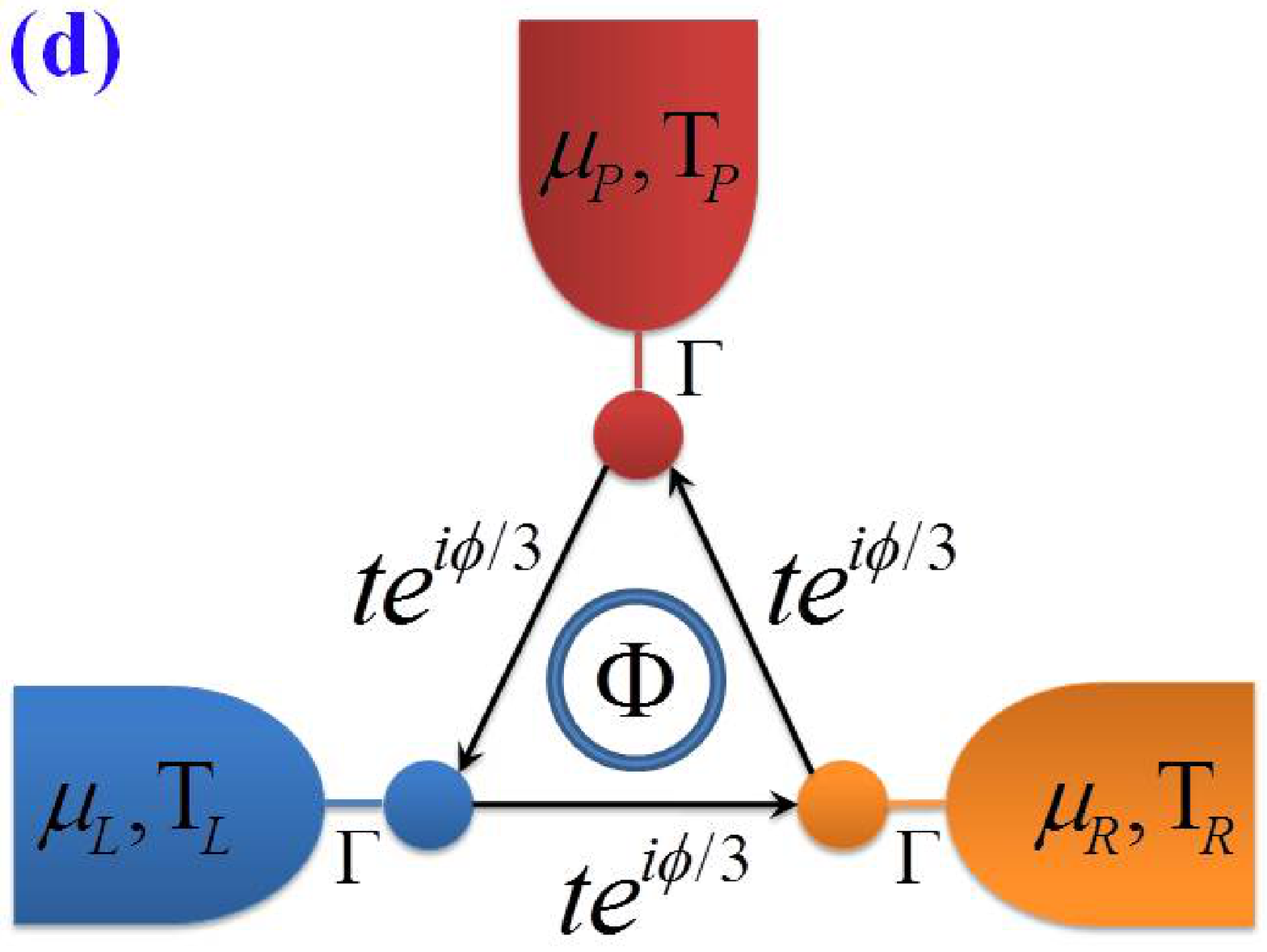}  
\caption{Efficiency statistics for TRS and TRB systems. (a) An example of the LDF
  $J(\eta)$ for TRS (blue, $r=1$) and TRB (green, $r=1.6$) cases with
  $\alpha=-0.3$ and $q=0.7$. (b) The least probable efficiency
  $\eta^\star$ for these two systems at different $\alpha$. (c) LDF
  $J(\eta)$ as a function of $\alpha$ and $\eta$ for a TRB system with
  $r=1.6$ and $q=0.7$. Note that the least probable efficiency (the
  maximum of the LDF) $\eta^\star$ shifts away from the Carnot 
  efficiency [$\eta_C\equiv 1$, labeled as the dashed line in (c)] for
  TRB systems as demonstrated in the panels (a), (b), and (c). 
  In panel (c) the minimum of $J(\eta)$
  for $\eta\gg 1$ with large negative $\alpha$ is associated with the
  reversed machine of which the actual efficiency is $1/\eta$, instead
  of $\eta$. 
(d) Our analysis is exemplified on a triple-QD thermoelectric device,
  see Fig. \ref{fig:3qd} and text.
%
}
\label{fig:jeta}
\end{center}
\end{figure}

The second law of thermodynamics requires
that \cite{casati,jiang} $\ov{{S}}_{tot}\ge 0$ and
hence $M_{11},M_{22}\ge 0$ and $M_{11}M_{22}\ge (M_{21}+M_{12})^2/4$,
i.e.,
\be
-2 \le q (1+r) \le 2 .\label{2nd-bound}
\ee
Equality is  attained only in the ``tight-coupling'' limit \cite{caplan} where the
average efficiency reaches its  upper bound. 
It has been proposed \cite{casati} that breaking time-reversal symmetry can open
the possibility of achieving Carnot efficiency at finite output power
and improving the efficiency at maximum output power to overcome the
Curzon-Ahlborn limit \cite{ca}. In contrast to the average efficiency
discussed in previous works
\cite{jiang,casati,trb2,seifert,trb3,nernst,edge}, the LDF $J(\eta)$ allows us to examine
the statistics of efficiency fluctuations. Particularly, we will show
that in TRB systems the tight-coupling limit becomes unfavourable as
the efficiency distribution becomes infinitely broad at {\em all times}. 

The LDF in TRB systems, Eq. (\ref{eq:jeta}), is a key expression in our work.
As we discuss below, its shape can be characterized by three quantities:
the average value of efficiency $\bar \eta$,
the least probable efficiency $\eta^\star$, and the width of the distribution around the average, $\sigma_{\eta}$.
In what follows we investigate these properties, particularly, under two (separate) experimentally relevant
conditions of maximum average efficiency and maximum average output power.

We begin with some general properties of the LDF.
First, the LDF at the Carnot efficiency $J(\eta_C)$
is {\it independent of affinities but it is solely determined by the response
coefficients}. It is also invariant under time-reversal operation, which turns
 $r
\to 1/r$ and $q \to qr$ \cite{casati,ft}. 
$J(\eta_C)$ can be suppressed by
breaking time-reversal symmetry, particularly in approaching the
tight-coupling limit. 
Second, in the TRS limit, $r=1$, Eq.~(\ref{eq:jeta}) goes back to results
obtained in Refs.~\cite{me1,me2,me3,me4}. In more general situations
we find that $0\le J(\eta)\le 1/4$ is guaranteed by the
thermodynamic bound (\ref{2nd-bound}) [see Supplementary Material].
Moreover, $J(\eta)$ has only one minimum and one maximum. While
the minimum $J(\ov{\eta})=0$ is reached at the average efficiency
$\ov{\eta}=-\alpha(\alpha+qr)/(\alpha q+1)$, the maximum value
$J(\eta^\star)=1/4$ is realized at the least probable efficiency
\be
\eta^\star = 1 + \frac{q (r -1) (1 + \alpha q + \alpha q r + \alpha^2 )}{q - q r - 2 \alpha + q^2 (1 + r) \alpha} .\label{eta-star}
\ee
In the TRS limit, the least likely efficiency is {\em always} identical to the Carnot
efficiency, $\eta^\star=\eta_C\equiv 1$ \cite{me1,me2,me3,me4}. For TRB
systems, in contrast, we find here that $\eta^\star$ {\em depends on} the parameters
$q$, $r$, and $\alpha$, see Fig.~\ref{fig:jeta}(a), (b), and (c).
%
%
%
We note that $\eta^\star$ diverges at $
\alpha_c = \frac{q (1 -  r)}{2 - q^2 - q^2 r}$.
For $|r|>1$, $\alpha_c$ produces a positive average efficiency and
output power, relevant for device operation. Besides,
$\eta^\star \xrightarrow{r\to \infty}\infty$
for all $\alpha$ and $q$. It
has been shown that this limit is achievable for a triple-QD
thermoelectric device [see Fig.~\ref{fig:jeta}(d) for schematic of the
device] when $M_{21}\to 0$ and $M_{12}\ne 0$ take place
simultaneously \cite{trb2}, see also Supplementary Materials.

The width of the distribution around the average efficiency, $\sigma_\eta$,  is another key 
characteristic of efficiency fluctuations.
Expanding $J(\eta)$ around
its minimum $\ov{\eta}$, one writes $J(\eta)\simeq
\frac{1}{2\sigma_{\eta}^2}(\eta - \ov{\eta})^2 + {\cal 
  O}((\eta-\ov{\eta})^3)$, to provide here
\be
\sigma_\eta = \frac{2 \sqrt{2}
  |\alpha| (1 - q^2 r) (1 + \alpha^2 + \alpha q + \alpha q r)}{(1 + \alpha q)^2 \sqrt{
    4 - q^2 (1 + r)^2}}. \label{sig}
\ee
We now proceed to describe the properties of $\bar \eta$, $\eta^\star$ and $\sigma_\eta$ under conditions for optimized (average)
operations.

\begin{figure}[]
\begin{center}
\includegraphics[height=3.74cm]{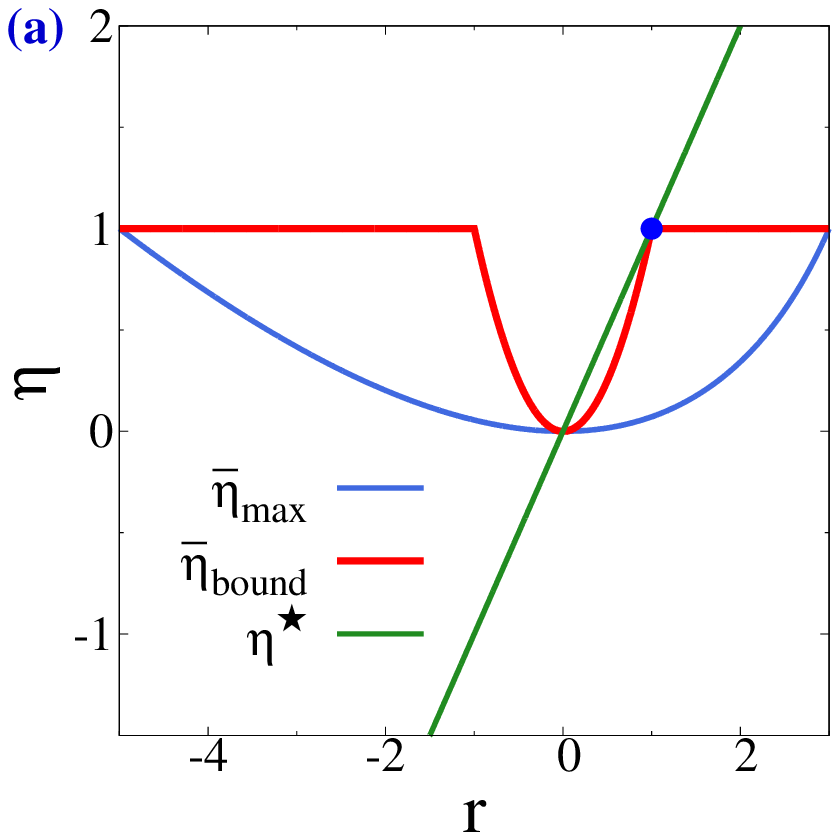}\includegraphics[height=3.8cm]{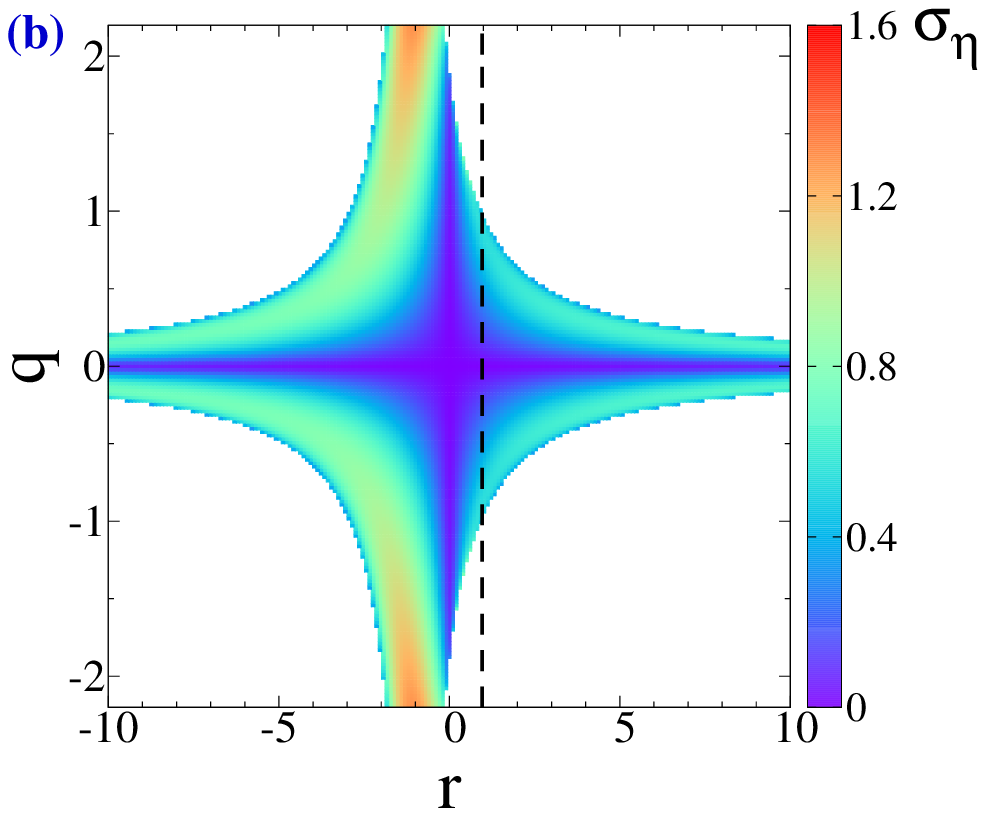}
\caption{Efficiency statistics under the maximum average efficiency
  condition. (a) Maximum average efficiency $\ov{\eta}_{max}$
  (blue curve), thermodynamic upper bound of the average
  efficiency $\ov{\eta}_{bound}$ (red curve), and the least probable
  efficiency $\eta^\star$ (green curve) for $q=0.5$. The blue dot
  represents the TRS point, $r=1$. (b) Width of efficiency distribution $\sigma_\eta$. The dashed line represents the TRS limit, 
  $r=1$. The white region is forbidden by the thermodynamic bound
  (\ref{2nd-bound}).}
\label{fig:width1}
\end{center}
\end{figure}

{\sl Efficiency fluctuations at maximum average efficiency}.---
We obtain the familiar form for the maximum average efficiency 
\be
\ov{\eta}=\ov{\eta}_{max} =r \frac{\sqrt{ZT+1}-1}{\sqrt{ZT+1}+1} = r\left(
\frac{1-\sqrt{1-q^2r}}{1+\sqrt{1-q^2r}} \right)
\ee
when $\alpha=-qr/(1+\sqrt{1-q^2r})$ \cite{jiang},
expressed in terms of the  figure of merit for energy conversion
$ZT=q^2r/(1-q^2r)$ \cite{casati,jiang}.
The thermodynamic upper bound of the average efficiency, reached at
the tight-coupling limit, $|q (1+r)|\to 2$, is \cite{casati,jiang}  
\be
\ov{\eta}_{bound}={\rm min}\{r^2,1\} .
\ee
One of our key results here is that under the maximum average
efficiency condition, the least probable efficiency (\ref{eta-star}) reduces to
($\eta_C\equiv 1$)
\be 
\eta^\star = r .
\ee
In the TRS limit this recovers recent findings that Carnot
efficiency is the least probable stochastic efficiency \cite{me1,me2}. The maximum
average efficiency, the upper bound of the average efficiency, and the
least probable efficiency are plotted in Fig.~\ref{fig:width1}(a) as a
function of $r$ for $q=0.5$, and we observe that
\begin{subequations}
\begin{align}
& \eta^\star \ge \ov{\eta}_{bound} \ge \ov{\eta}_{max}, \quad \forall r\ge 0, \\
& \eta^\star < \ov{\eta}_{max} \le \ov{\eta}_{bound}, \quad \forall
r<0 .
\end{align}
\end{subequations}
The least probable efficiency $\eta^\star$ coincides with
$\ov{\eta}_{bound}$ only in the TRS limit, $r=1$,  or for the TRB case with $r=0$.

The width parameter $\sigma_\eta$ is plotted in
Fig.~\ref{fig:width1}(b) where the white region in the figure is 
forbidden by the second law of thermodynamics according to
(\ref{2nd-bound}). It is small
when $q$ or $r$ are  small, 
corresponding to the ``weak-coupling limit'' when the average
efficiency is small. Physically, this can be understood as that two
nearly uncoupled currents are very unlikely to have collective
fluctuations which result in a considerably large efficiency. 



Approaching the tight-coupling limit $|q(1+r)|\to 2$ [which cannot be depicted in Fig.~\ref{fig:width1}(b)], we find
that $\sigma_{\eta} \to \infty$ for $0<|r|<1$ and $r=-1$, while $\sigma_{\eta} \to 0$ for other regimes.
In the TRS limit, $r=1$, our results agree with Ref.~\cite{me3}.
The singular behavior of $\sigma_\eta$ in this limit can be understood by noticing
that the denominator of Eq.~(\ref{sig}) vanishes in the tight-coupling
limit whereas the numerator is proportional to the (average) total
entropy production rate. It has been shown \cite{casati,jiang} that in
the tight-coupling limit, for $|r|<1$, the maximum average efficiency
is attained at {\em finite} (average) total entropy production rate
(thus the upper bound efficiency is {\em less than} 100\%). In
contrast, for $|r|\ge 1$, the maximum average efficiency is reached
when the (average) total entropy production rate is zero (hence the
upper bound efficiency is 100\%). When $\ov{S}_{tot}=0$ the unscaled
LDF ${\cal G}(\eta)=\ov{S}_{tot}J(\eta)$ is always zero. Therefore, in
the tight-coupling limit the distribution of stochastic efficiency is
infinitely broad for $|r|\ge 1$ as well. The only exception is the TRS
limit, $r=1$, where the width of efficiency distribution becomes
zero\cite{me3}.

Direct examination of Eq.~(\ref{eq:jeta}) shows that in the
tight-coupling limit $J(\eta_C)=0$ (except for $r=1$, the TRS limit).
$J(\eta)$ thus vanishes whenever the average entropy production rate
$\ov{S}_{tot}$ is nonzero. However, when $\ov{S}_{tot}=0$, $J(\eta)$
is ill-defined, nevertheless the unscaled LDF ${\cal
  G}(\eta)=\ov{S}_{tot}J(\eta)$ is always zero. In addition, for $r=0$
and $\alpha=0$, $J(\eta)$ is constant for all $\eta$. Therefore, the 
distribution of efficiency is infinitely broad for {\em any} TRB energy
transducer in the tight-coupling limit.

An example that may help to understand the anomaly in the
tight-coupling limit is the integer quantum Hall effect. For example,
at filling factor $\nu=1$, electrical transport in the quantum
Hall insulator is described by
\be
\left( \begin{array}{cccc} j_x\\ j_y \end{array}\right) =
\frac{e^2}{h} \left( \begin{array}{cccc} 0 & 1 \\ -1 &
    0 \end{array} \right) \left( \begin{array}{cccc} {\cal E}_x\\
    {\cal E}_y \end{array}\right) 
\ee
where $j_{x/y}$ and ${\cal E}_{x/y}$ are the electrical currents and
fields along the $x/y$ direction, respectively. Although the quantum
Hall insulator does not conduct electron longitudinally, charge can be
conducted via the Berry phase effect, or more physically through the
chiral edge states. In this way, the system converts electrical energy
in the $x$ direction $W_x=j_x{\cal E}_x=\frac{e^2}{h}{\cal E}_x{\cal
  E}_y$ to electrical energy in the $y$ direction $W_y=j_y{\cal
  E}_y=-\frac{e^2}{h}{\cal E}_x{\cal E}_y$ (and vice versa). The
macroscopic efficiency $\eta=-W_y/W_x$ is always 100\%, and the
output power $-W_y$ is finite for nonzero (positive) ${\cal E}_x{\cal
  E}_y$.

However, in the Gaussian description, the distributions of the
electrical currents $j_x$ and $j_y$ are singular because the transport
is completely dissipationless. This also leads to singular distributions
of the output power and efficiency. Casting into our parameters, the
integer quantum Hall systems have $q= \infty$ and
$r=-1$. Nevertheless, in a finite system the longitudinal conductance
is not vanishing (i.e., $q$ is finite). A further examination of current
noises in a Hall measurement needs more careful
treatment \cite{qhe}, which is beyond the scope of this work. We
speculate that close to the tight-coupling limit, the Gaussian
approximation is inadequate to describe fluctuations in the system.
However, energy efficiency is much easier to measure in this
system since the measurement of stochastic electric currents in
mesoscopic systems is a rather mature technology\cite{qhe}. It is
much easier to measure stochastic efficiency distribution in this
time-reversal symmetry broken system than in other known solid-state
systems.

\begin{figure}[]
\begin{center}
\includegraphics[height=3.8cm]{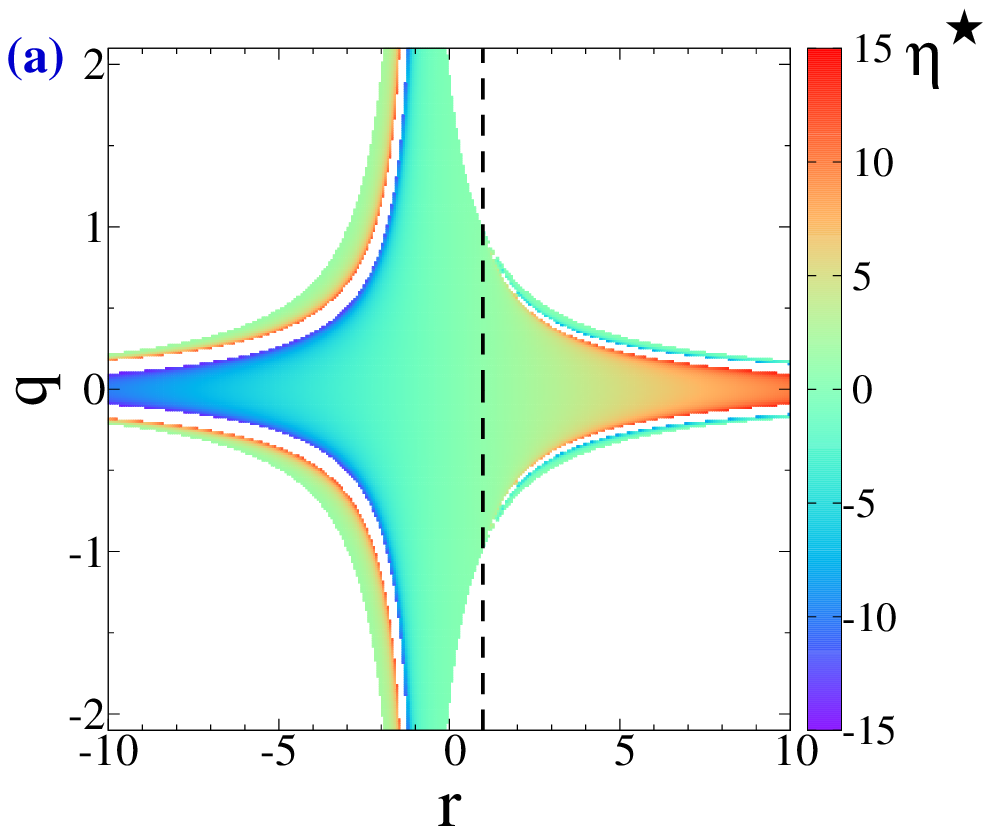}\includegraphics[height=3.8cm]{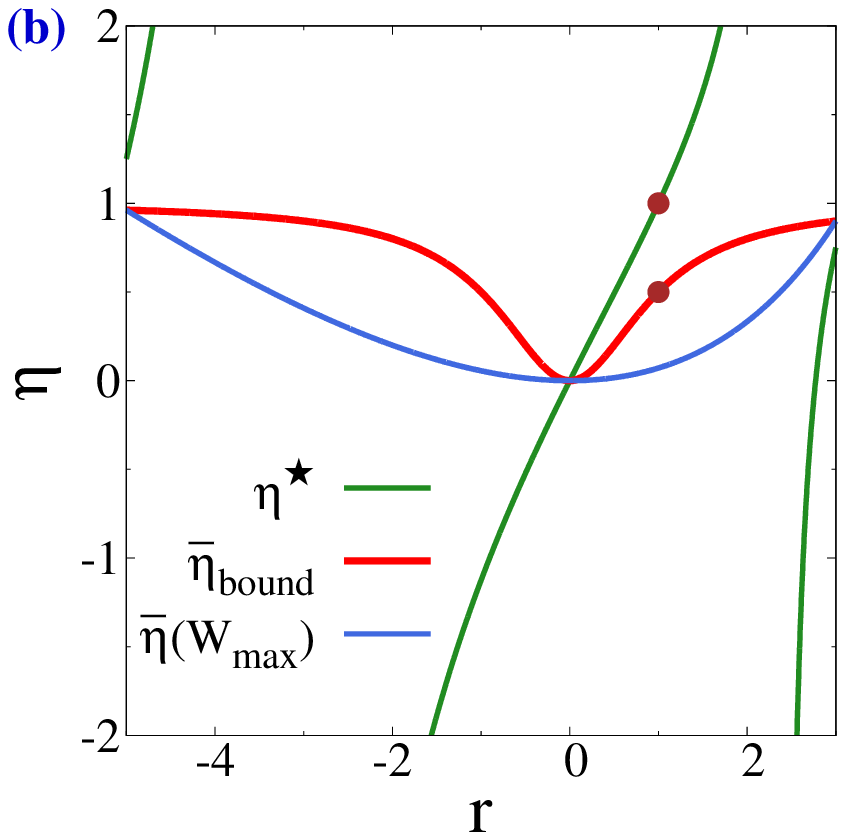}
\caption{Efficiency statistics under the maximum average power
  condition. (a) Least likely efficiency $\eta^\star$ as a
  function of $r$ and $q$. The dashed line represents the TRS limit
  $r=1$. The white region outside is forbidden by the thermodynamic bound
  (\ref{2nd-bound}). In contrast, the white region inside the color
  graphics region is due to divergences of $\eta^\ast$ at special
  $r$-$q$ lines given by $ r = (-1 \pm \sqrt {1 + 8/q^2})/2$.  
  (b) Least probable efficiency $\eta^\star$,  average
  efficiency $\ov{\eta}(W_{max})$  and its upper bound
  $\bar\eta_{bound}(W_{max})$ at $q=0.5$ as a function of $r$. The 
  dots identify values in the TRS limit, $r=1$.}
\label{fig:peak2}
\end{center}
\end{figure}

{\sl Efficiency fluctuations at maximum average output power}.---
We now turn our attention to another highly pursued situation, the maximum average output power
condition \cite{martinez} arrived at $\alpha=-qr/2$. Under
this condition the average efficiency is \cite{casati,jiang} 
\be
\ov{\eta}(W_{max}) = \frac{rZT}{2(ZT+2)} = \frac{q^2r^2}{4-2q^2r} ,
\ee
with the thermodynamic upper bound \cite{casati,jiang} 
\be
\ov{\eta}_{bound}(W_{max}) = \frac{r^2}{1+r^2} ,
\ee 
reached in the tight-coupling limit. Note that for $r^2>1$ the
efficiency at maximum power can be larger than 50\% (the
value of the Curzon-Ahlborn efficiency in the linear-response regime).
The least likely efficiency is found to be
\be
\eta^\star = r \left ( \frac { 4 - 3 q^2 r  - q^2 r^2 } {4 - 
    2 q^2 r - 2 q^2 r^2 } \right) . \label{empstar}
\ee
Calculations in Fig.~\ref{fig:peak2} indicate
that there are lines of singularities for $\eta^\star$, varying with
$q$ and $r$. These singularity lines appear at $ r = (-1 \pm
\sqrt {1 + 8/q^2})/2$, and they reach to the thermodynamic bound in
the $r$-$q$ plane at $r=1$ and $q=\pm 1$. The width of efficiency
distribution under maximum average output power condition also shows
the same singular behavior as that under the the maximum average
efficiency condition. In the Supplementary Material we explore
$J(\eta)$ under different situations and manifest its rich features.

{\sl TRB thermoelectric transport in three-terminal systems}.---
We exemplify our analysis within a {mesoscopic} triple-QD thermoelectric device,
see Fig. \ref{fig:jeta}(d).
The affinities for a two-terminal thermoelectric device are {the
  electrochemical potential} $A_1=\frac{\mu_L-\mu_R}{eT_R}$ and
{the temperature difference} $A_2=\frac{1}{T_R}-\frac{1}{T_L}$,
where $e$ is the electronic charge,
 $T_{L,R}$ and $\mu_{L,R}$ denote
the temperatures and chemical potentials in the left ($L$) and right
($R$) electronic reservoirs. 
In order to receive $M_{12}\neq M_{21}$,
we introduce a third (probe $P$) terminal and employ the constraints
that the average thermal and electrical currents flowing out of the
probe terminal are zero \cite{2t,seifert}. These conditions set the temperature 
$T_P$ and chemical potential $\mu_P$ in the probe.
Each QD is
coupled through elastic tunneling to the nearby reservoir
thus we employ the
indices $1/2/3$ to identify the leads $L/R/P$, respectively.
Hopping
between QDs are affected by the magnetic flux $\Phi$ piercing
through at the center with $\phi=2\pi\Phi/\Phi_0$ ($\Phi_0$ is flux
quantum). 
The system is described by the Hamiltonian 
$\hat{H}=\hat{H}_{qd}+\hat{H}_{lead}+\hat{H}_{tun}$ where
$\hat{H}_{qd}=\sum_{i=1,2,3} E_i d_i^\dagger d_i + (t e^{i\phi/3}
d_{i+1}^\dagger d_i + {\rm H.c.})$,
$\hat{H}_{lead}=\sum_{i=1,2,3}\sum_k\vep_{k} c^\dagger_{ik} c_{ik}$,
and $\hat{H}_{tun}=\sum_{i,k}V_{k}d^\dagger_ic_{ik}+{\rm H.c.}$. 
 
\begin{figure}[]
 \begin{center}
   \includegraphics[height=3.8cm]{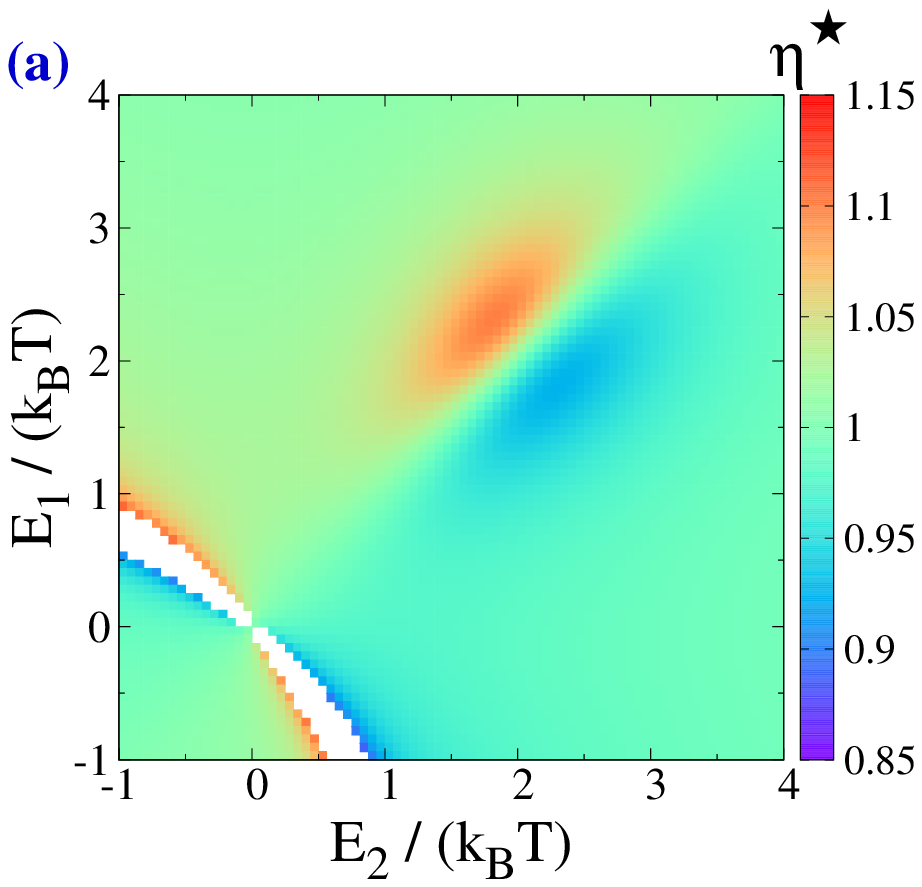}\includegraphics[height=3.8cm]{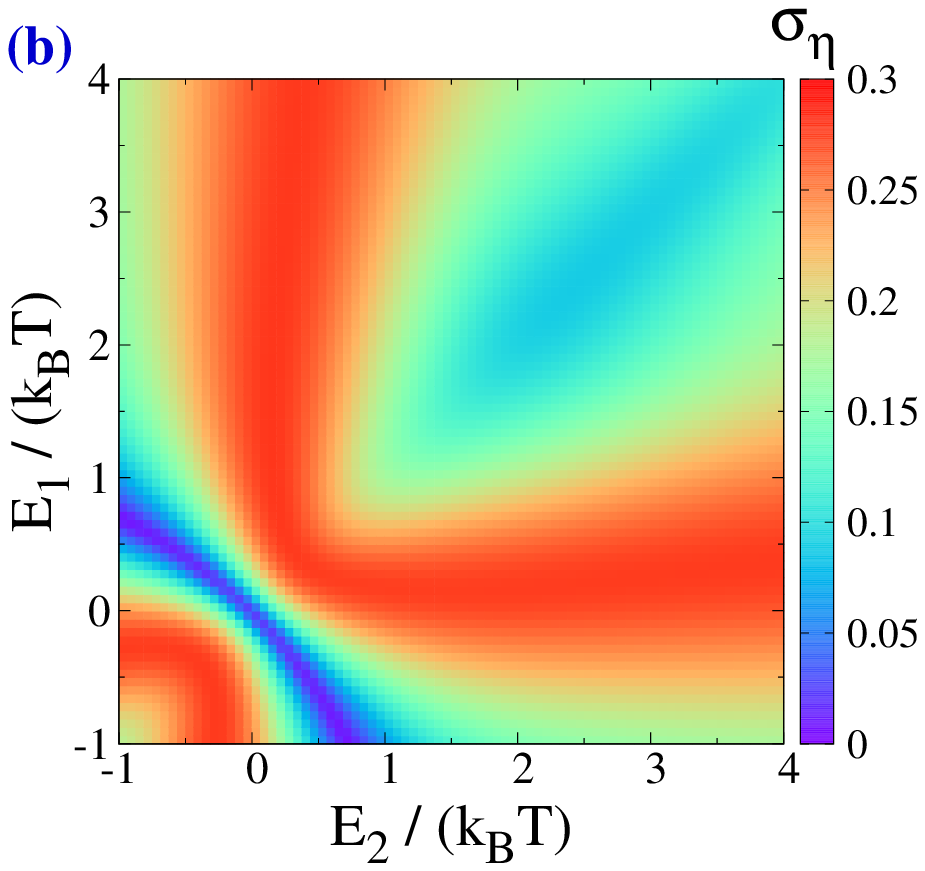}
 \caption{Triple-QD thermoelectric systems at the maximum average output
   power condition: (a) Least
   probable efficiency $\eta^\star$ and (b) width of efficiency distribution $\sigma_\eta$.
The QD connected to the probe terminal
  is set at $E_3=2$. $\phi=\pi/2$, $\Gamma=0.5$, and $t=-0.2$. 
 The white region in (a) depicts very
   large or very small (negative) $\eta^\star$ values which are not properly
   incorporated into the figure.}
 \label{fig:3qd}
 \end{center}
 \end{figure}

This noninteracting model has been analyzed thoroughly in
Ref.~\cite{trb2} using the Landauer-B\"uttiker approach, 
to study transport properties and the average efficiency.
We recall here that the transmission function is given by
$\hat{S}(\ome)=-\hat{1}+i\Gamma \hat{G}^r(\ome)$ where
$\hat{G}^r(\ome)=[(\ome + i\Gamma/2 )\hat{1} - \hat{H}_{qd}]$ is the
retarded Green's function of the quantum dots and the damping rate
$\Gamma=2\pi\sum_{k}|V_{ik}|^2\delta(\ome-\vep_{ik})$ is assumed to be
a constant (independent of energy) for all three leads.
Using $\phi=\pi/2$, $\Gamma=0.5$ and $t=-0.2$ (the
equilibrium chemical potential is set at zero, the energy unit is $k_BT$), we
calculate transport coefficients and substitute them into
Eq.~(\ref{eq:jeta}) to obtain the LDF of stochastic efficiency.
We then calculate the least probable efficiency $\eta^\star$ and the
width of distribution $\sigma_\eta$ under the maximum average output
power condition, see Fig.~\ref{fig:3qd}. We find that around
$E_1+E_2=0$, $|\eta^\star|$ becomes very large. The underlying physics 
is that when $E_1+E_2$ is close to zero, the transport coefficient
$|M_{21}|$ can become very small while $|M_{12}|$ is still finite, to yield
a very large $|r| $\cite{trb2}. Accordingly, the least probable
efficiency $|\eta^\star|$ becomes very large as we showed in Fig. \ref{fig:peak2}(b). In addition, at this
special region the width 
$\sigma_\eta$ becomes very small, in accordance with Eq.~(\ref{sig}),
for $\alpha=-qr/2$ ($|qr|\ll 1$, $q\to 0$). These calculations, based on the Gaussian
approximation, agree with a careful full-counting statistics analysis
with vertex corrections \cite{utsumi}, carried out in the
linear-response regime [see Supplementary Material].

{\sl Future perspectives.}--- Our analysis lays the foundation for
studies of efficiency statistics beyond the linear-response
approximation, as well as to experimental works via, e.g., mesoscopic
quantum Hall systems\cite{edge} or TRB thermoelectric engines.


{\sl Acknowledgments.}--- We thank Matteo Polettini and Massimiliano
Esposito for helpful discussions. J.H.J acknowledges support from
start-up funding of Soochow University.  He also thanks Yoseph Imry and Karen Michaeli for
helpful discussions, and the Weizmann Institute of Science for hospitality.
D.S and B.K.A acknowledge support from an NSERC
Discovery Grant, the Canada Research Chair program,
and the CQIQC at the University of Toronto.

\newpage

\begin{widetext}

\section{Supplementary Materials}

\section{I. LDF in TRB systems and its properties}

\subsection{IA. Derivation of the LDF for TRB systems}

The LDF of efficiency fluctuations was derived in Ref. \cite{me3},
specific to TRS systems. Here we extend this work and obtain the LDF for TRB systems.
We begin by introducing the probability distribution function (PDF)
of the stochastic currents $I_i$ ($i=1,2$) 
\be
P_t(\vec{I})=
\frac{t\sqrt{\det((\hat{M}^{-1})_{sym})}}{4\pi} 
\exp(-\frac{t}{4}\delta\vec{I}^T\cdot\hat{M}^{-1}\cdot\delta
\vec{I}) .
\ee
Replacing the stochastic currents with stochastic entropy production
rates $S_i=I_iA_i$ ($i=1,2$) at given affinities $A_1$ and $A_2$, one
finds the PDF of the entropy production\cite{me2,me3}
\be
P_t(S_1, S_2) = \frac{t \sqrt{\det((\hat{C}^{-1})_{sym})}}{4\pi
  \ov{{S}}_{tot}}\exp\left[-\frac{t}{4\ov{{S}}_{tot}}\delta
  \vec{S}^T\cdot \hat{C}^{-1} \cdot \delta \vec{S} \right] ,
\ee
where $\delta \vec{S}=\vec{S} -\vec{\ov{{S}}}$, $\vec{S}=(S_1,S_2)^T$ is
the stochastic entropy production, $\vec{\ov{{S}}}=(\ov{{S}}_1,\ov{{S}}_2)$
is the averaged (macroscopic) entropy production rate, and
$\det((\hat{C}^{-1})_{sym})$ is the determinant of the symmetric part of
the inverse of matrix $\hat{C}$. The macroscopic
total entropy production rate is $\ov{{S}}_{tot}=\ov{{S}}_1+\ov{{S}}_2$. 
Here, $C_{ij}=A_iM_{ij}A_j/\ov{{S}}_{tot}$ ($i,j=1,2$). The PDF of
stochastic efficiency is
\be
P_t(\eta) = \int d S_1 d S_2 \delta (\eta + \frac{S_1}{S_2} )
P_t(S_1, S_2) = \int dS_2 |S_2| P_t(-\eta S_2, S_2) .
\ee
A direct calculation yields an expression 
\be
P_t(-\eta S_2, S_2) = \frac{t \sqrt{\det((\hat{C}^{-1})_{sym})}}{4\pi \ov{{S}}_{tot} }\exp\left[-\frac{t}{4\ov{{S}}_{tot}}[a(\eta)S_2^2 +
  2 b(\eta) S_2 + c]\right] ,
\ee 
with the coefficients
\begin{subequations}
\begin{align}
a(\eta) &= (C_{11} + \eta ( C_{12} + C_{21} ) + \eta^2 C_{22} )/{\rm
  det}(\hat{C}) , \quad \quad c  = \ov{{S}}_{tot}^2 ,\\
b(\eta) &= \{ \ov{{S}}_1 (C_{12}+C_{21}) - 2\ov{{S}}_2 C_{11} + \eta
[2C_{22}\ov{{S}}_1 - (C_{12}+C_{21})\ov{{S}}_2]\}/[2\det(\hat{C})] . 
\end{align}
\end{subequations}
The coefficients $C_{ij}$ sum up to unity $\sum_{ij}C_{ij}=1$.
$\det(\hat{C})=C_{11}C_{22}-C_{12}C_{21}$ is the determinant of
$\hat{C}$. Note that $C_{12}\ne C_{21}$ because of time-reversal
symmetry breaking. The full probability
distribution of the stochastic efficiency is now found to be
\be
P_t(\eta) = \frac{\sqrt{\det((\hat{C}^{-1})_{sym})}\exp[-t\ov{{S}}_{tot}/4]}{\pi a(\eta)
  }\left[1+\sqrt{\pi t \ov{{S}}_{tot}} h(\eta)
  \exp[t \ov{{S}}_{tot} h^2(\eta)] {\rm erf}(\sqrt{t
  \ov{{S}}_{tot}} h(\eta))\right] ,
\ee
with ${\rm erf}(x)=\frac{2}{\sqrt{\pi}}\int_0^xe^{-t^2}dt$ being the
error function and 
\be
h(\eta) = \frac{-b(\eta)}{2\ov{{S}}_{tot}\sqrt{a(\eta)}} .
\ee
The large deviation function of stochastic efficiency is obtained from
\be
J(\eta) \equiv - \frac{\lim_{t\to
    \infty}\ln[P_t(\eta)]}{ t \ov{{S}}_{tot}} = \frac{1}{4} - h^2(\eta).
\ee
By substituting the following parametrization,
\be
q \equiv \frac{M_{21}}{\sqrt{M_{22}M_{11}}},\quad r \equiv \frac{M_{12}}{M_{21}},
\quad \alpha \equiv \frac{A_1\sqrt{M_{11}}}{A_2\sqrt{M_{22}}} ,
\ee
we find that
\begin{subequations}
\begin{align}
a(\eta) & = \frac{ ( 1 + \alpha q + \alpha r q +
\alpha^2 ) (\alpha^2 + \alpha q (1 + r) \eta + \eta^2)}{
 \alpha^2 (1 - q^2 r)} ,\\
b(\eta) & = M_{22} A_2^2 \frac{(  1 + \alpha q + \alpha r q +
\alpha^2 ) (\alpha^2 q (r -1) + 
    \alpha (q^2 (1 + r) (r - \eta) + 2 (\eta-1)) + 
    q ( r -1) \eta)}{2 \alpha (1 - q^2 r)}  ,\\
\ov{{S}}_{tot} & = M_{22} A_2^2 ( 1 + \alpha q + \alpha r q +
\alpha^2 ), \\
J(\eta) & = \frac { \left [4 - 
      q^2 (1 + r)^2 \right] \left ( \alpha^2 + \eta +  \alpha q r + \alpha q \eta \right)^2} {16\left (1 - 
     q^2 r \right) (1 + \alpha^2 + \alpha q + \alpha q r)\left ( \alpha^2 +  \alpha q\eta +  \alpha q r
     \eta + \eta^2 \right)} .\label{eq:jeta-ap}
\end{align}
\end{subequations}

\subsection{IB. Thermodynamic bounds on the LDF}

We prove here that the inequalities $0\le J(\eta)\le 1/4$ are guaranteed by the
thermodynamic bound 
\be
|q(1+r)|\le 2 .
\ee
First, $4(1 - q^2 r)\ge 4 - q^2 (1 + r)^2\ge 0$, which guarantees the
positive semi-definiteness of the prefactors of the numerator and
denominator in Eq.~(\ref{eq:jeta-ap}). Second, $1 + \alpha^2 + \alpha q + \alpha q r=\left(\alpha + 
  \frac{q(1+r)}{2}\right)^2+\left(1-\frac{q^2(1+r)^2}{4}\right)\ge
0$. This is consistent with the fact that this term originates from
the average total entropy production rate. The last term in the
denominator is also not less than zero, since  $\alpha^2 + \alpha q\eta + \alpha q\eta r +
\eta^2=\left(\alpha+\eta\frac{q(1+r)}{2}\right)^2+\eta^2\left(1-\frac{q^2(1+r)^2}{4}\right)\ge
0$. Therefore $J(\eta)$ is guaranteed to be greater than zero.
Using exactly the same arguments, one can show that
\be
\frac{1}{4}-J(\eta) = \frac{
[\alpha^2 q (r -1) + \alpha (q^2 (1 + r) (r - \eta) + 2 (\eta -1 )) + 
  q (r -1) \eta]^2 }{16\left (1-q^2r\right) (1 + \alpha^2 + \alpha q + \alpha q r)\left ( \alpha^2 +  \alpha q\eta +  \alpha q r
     \eta + \eta^2 \right)} \ge 0 .
\ee
Therefore, $0\le J(\eta)\le 1/4$. From the above we find that
$J(\eta)=1/4$ is reached only at
\be
\eta^\star = 1 + \frac{q (r -1) (1 + \alpha q + \alpha q r + \alpha^2 )}{q - q r - 2 \alpha + q^2 (1 + r) \alpha} .
\ee
We also find that there is only one minimum of $J(\eta)$ which is the
macroscopic efficiency $\ov{\eta}$, and only one maximum of $J(\eta)$
which is precisely the least probable efficiency $\eta^\star$ given above.
This is confirmed by solving the extremum equation $\partial_{\eta}
J(\eta) =0$ where we find only two solutions: one is $\ov{\eta}$, the
other is $\eta^\star$. This property determines the basic-generic shape of
the LDF curve.

\subsection{IC. $J(\eta)$ under time-reversal operation}
In the main text we defined the following parameters
\begin{equation}
r = \frac{M_{12}}{M_{21}}, \quad q = \frac{M_{21}}{\sqrt{M_{22} M_{11}}}.
\end{equation}
Under time-reversal operation, $\phi \rightarrow -\phi$, the above
parameters transform as follows
\begin{equation}
r(-\phi) \rightarrow \frac{1}{r},\quad q(-\phi) \rightarrow q r,
\end{equation}
where we have used Onsager's reciprocity relation
$M_{12}(\phi)=M_{21}(-\phi)$. Denoting the LDF of efficiency for the 
reversed magnetic field ($\phi \to -\phi$) by $\tilde{J}(\eta)$, we
obtain that
\begin{equation}
J(\eta)-\tilde{J}(\eta)= \frac{J(\eta_C)}{1+ \alpha^2 + \alpha q r +
  \alpha q} \frac{\alpha^2 q^2 (r^2 -1) (1-\eta^2) + 2 (\alpha^2 +
  \eta) \alpha q (r-1) (1-\eta)}{\alpha^2 + \alpha q \eta + \alpha q r
  \eta + \eta^2} .
\end{equation}
We find that for TRB systems Carnot efficiency ($\eta_C=1$) appears as a
special point where the distributions become invariant under
time-reversal operation, i.e., $J(\eta_C) = \tilde{J}(\eta_C)$.
For TRS systems ($r=1$) this equality trivially holds for all values of efficiency.

\subsection{ID. $J(\eta)$ at various limits}

In this section we illustrate the rich behavior of $J(\eta)$ at
various limits: (i) weak coupling limit $r\to 0$
or $q\to 0$; (ii) tight-coupling limit with maximum macroscopic
efficiency for $0<|r|<1$, $|r|>1$, and $r=\pm 1$; (iii) tight
coupling limit with maximum macroscopic output power for $|r|\ne 0, 1$,
and for $r=\pm 1$. Results are plotted in Fig.~\ref{fig:lim}.

Fig.~\ref{fig:lim}(a) shows that in the weak-coupling limit
($r\to 0$), the LDF experiences a sharp transition from the minimum
$\ov{\eta}\to 0$ to the maximum $\eta^\star\to 0$. Therefore, $J(\eta)$
behaves like a derivative of the Dirac delta function. In
contrast, when $q\to 0$ but $r$ is finite, Fig.~\ref{fig:lim} (b) shows
that $J(\eta)$ develops an infinitely narrow dip near $\eta=0$, i.e.,
it behaves like the Dirac delta function itself. Note that in the weak
coupling regime with $q\to 0$ and/or $r\to 0$, the affinity parameter
$\alpha\to 0$. As a result, the behavior of $J(\eta)$ is very similar either under the maximum
macroscopic efficiency condition or the maximum output power condition as both the efficiency and the output power go to 0.

In the tight-coupling limit, the behavior is quite different under
those two conditions. We first examine the maximum macroscopic
efficiency condition. Fig.~\ref{fig:lim}(c) shows that for $0<|r|<1$, the
width of efficiency distribution  $\sigma_\eta$ tends to infinity,
while the maximum value of $J(\eta)$ at $\eta^\star=r$ develops a very
sharp peak. In contrast, for $|r|>1$, as shown in
Fig.~\ref{fig:lim}(e), the width of efficiency distribution  approaches
zero, while the width at the least probable efficiency becomes
infinite. The $r=1$ situation, see Fig.~\ref{fig:lim}(d), demonstrates a sharp
transition from the minimum value to the maximum point, resembling the behavior of the
derivative of the Dirac delta function. For $r=-1$, the tight
coupling limit, $|q(1+r)|\to 2$ is pushed to $q\to \infty$. Therefore, in this situation
for any finite $q$ the behavior of $J(\eta)$ shows
a regular behavior. Nevertheless, as shown in Fig.~\ref{fig:lim}(f), the
distribution of $\eta$ is quite broad for $r=-1$. This case is
relevant to recent studies on ``chiral 
thermoelectrics'' (e.g., Nernst engines) where, however, a much
stronger bound on $q$ was obtained \cite{trb3,trb4}.

Under the maximum macroscopic output power condition, the behavior of
$J(\eta)$ for $r\to 0$, $q\to 0$ cases is identical to that observed in 
Fig.~\ref{fig:lim}(a) and (b). The tight-coupling limit with $0<|r|<1$
also shows features similar to Fig.~\ref{fig:lim}(c). However, when $r=1$ in the
tight-coupling limit, see Fig.~\ref{fig:lim}(g), the width around the
macroscopic efficiency ($\ov{\eta}=0.5$) approaches zero while the width
around the maximum value of $J(\eta)$, at $\eta^\ast=1$, becomes
infinite.  Fig.~\ref{fig:lim}(h) focuses on the $|r|>1$ case for which the
width of efficiency distribution  tends to
infinity, while the maximum of $J(\eta)$ becomes a sharp
peak. Fig~\ref{fig:lim}(i) shows that at $r=-1$, $J(\eta)$ becomes
an extremely broad distribution.

\begin{figure}[]
  \begin{center}
    \includegraphics[height=3.5cm]{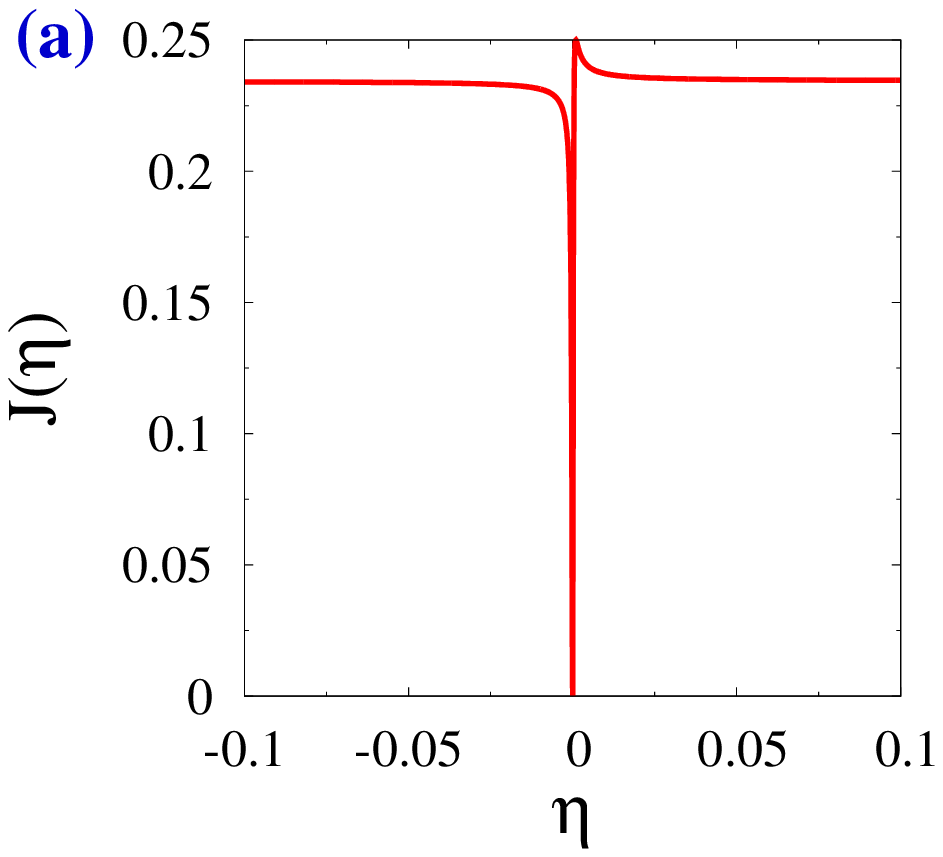}\includegraphics[height=3.5cm]{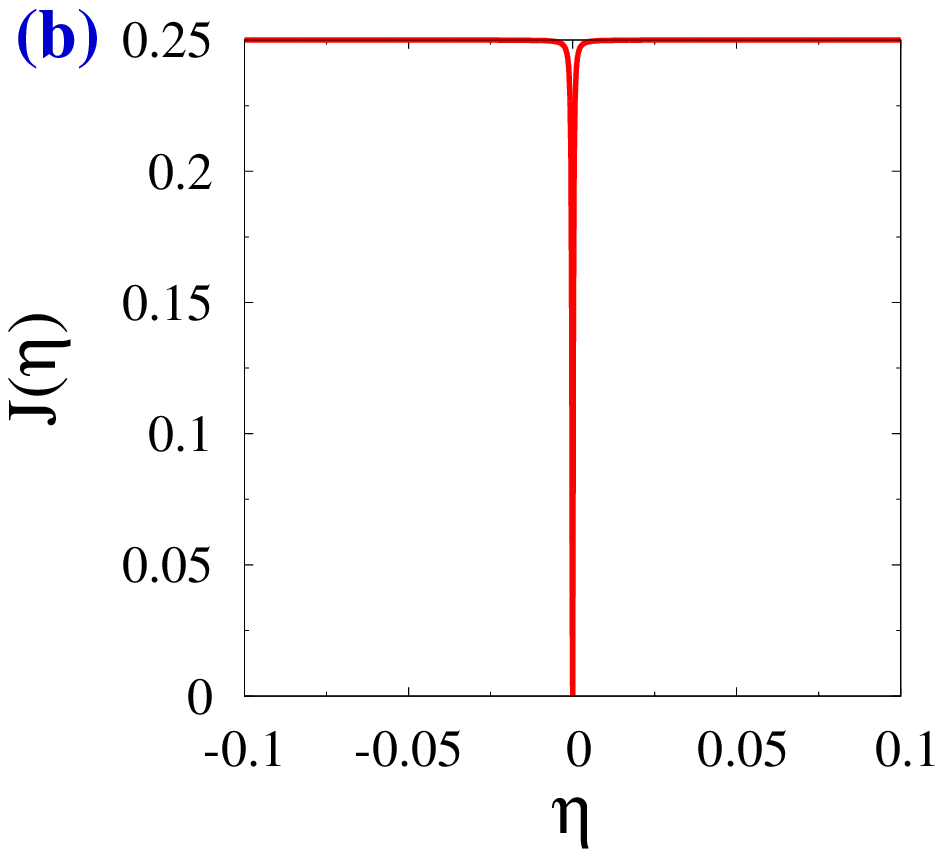}\includegraphics[height=3.5cm]{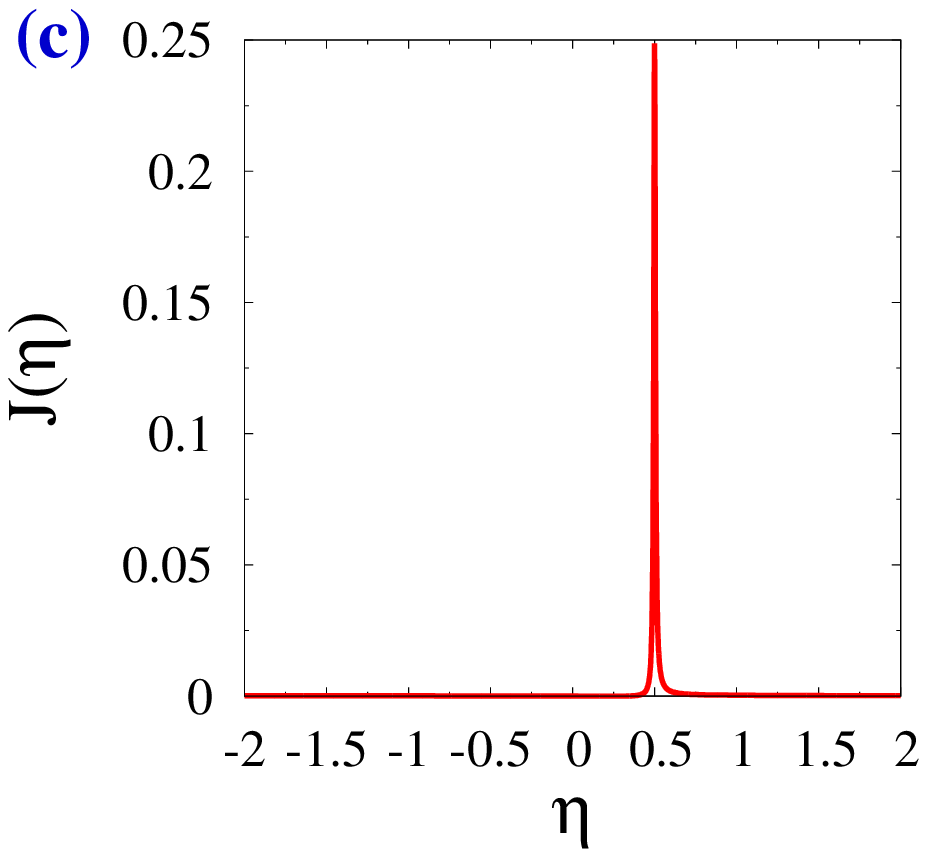}
    \includegraphics[height=3.5cm]{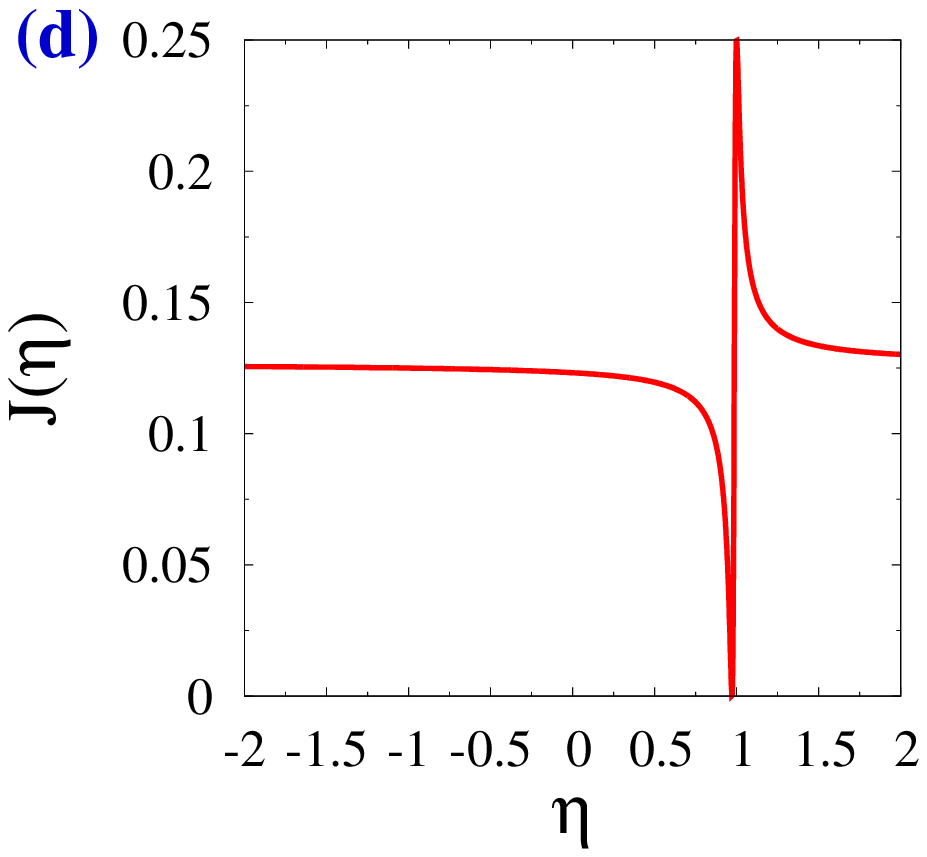}\includegraphics[height=3.5cm]{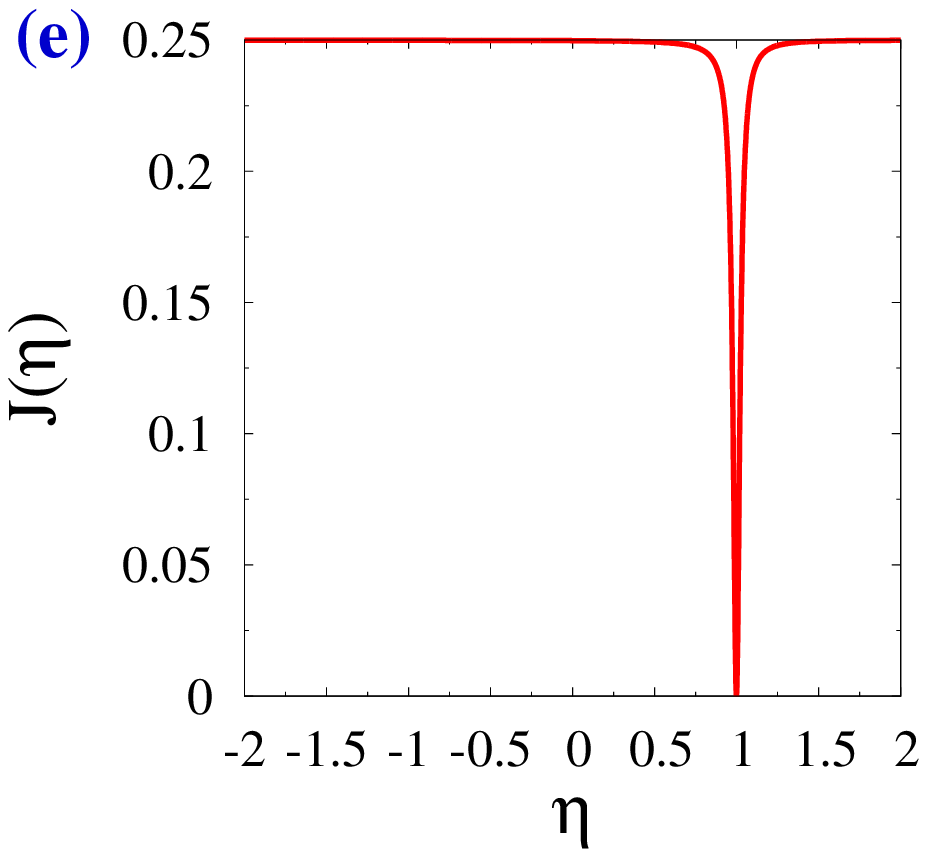}\includegraphics[height=3.5cm]{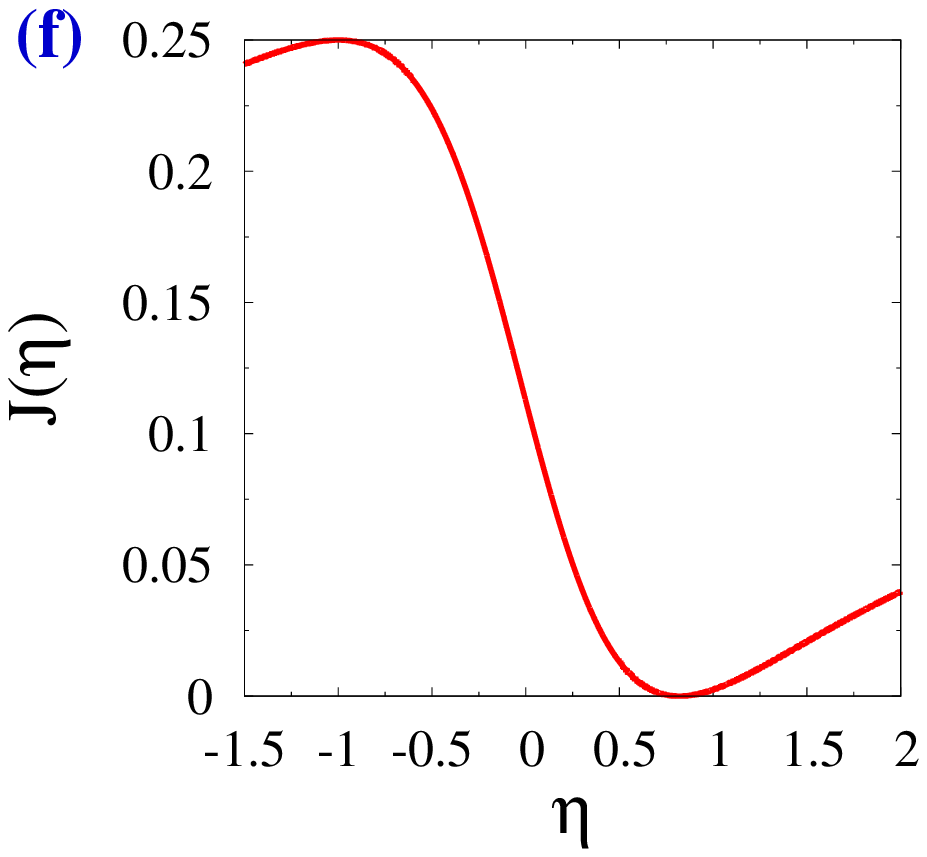}
    \includegraphics[height=3.5cm]{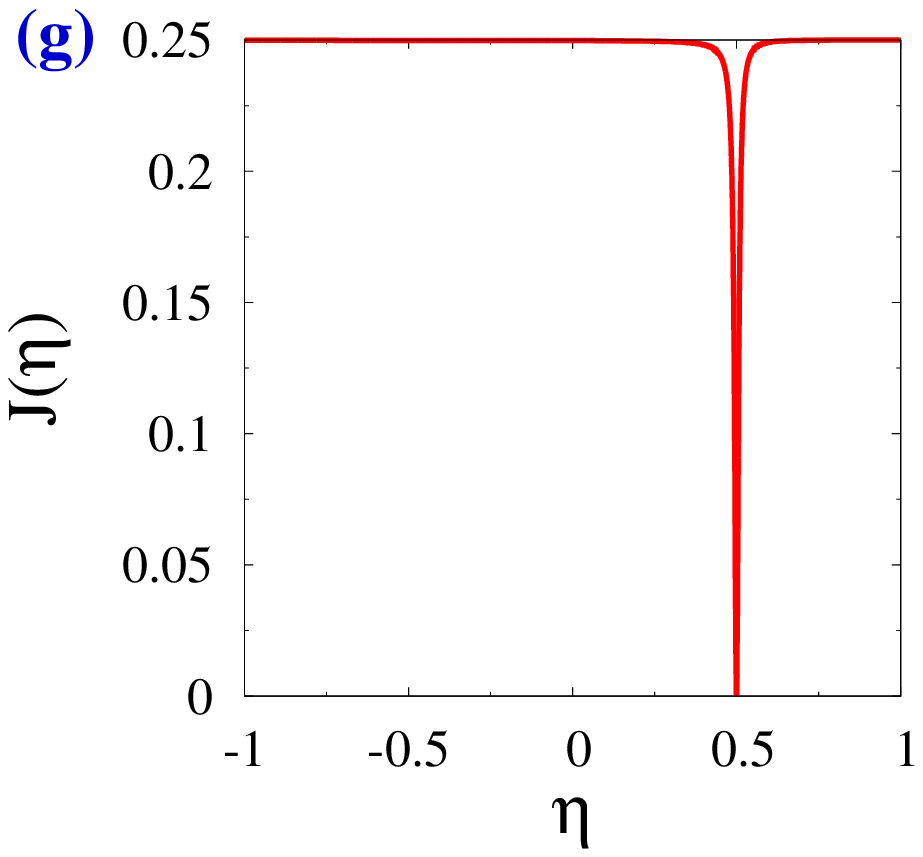}\includegraphics[height=3.5cm]{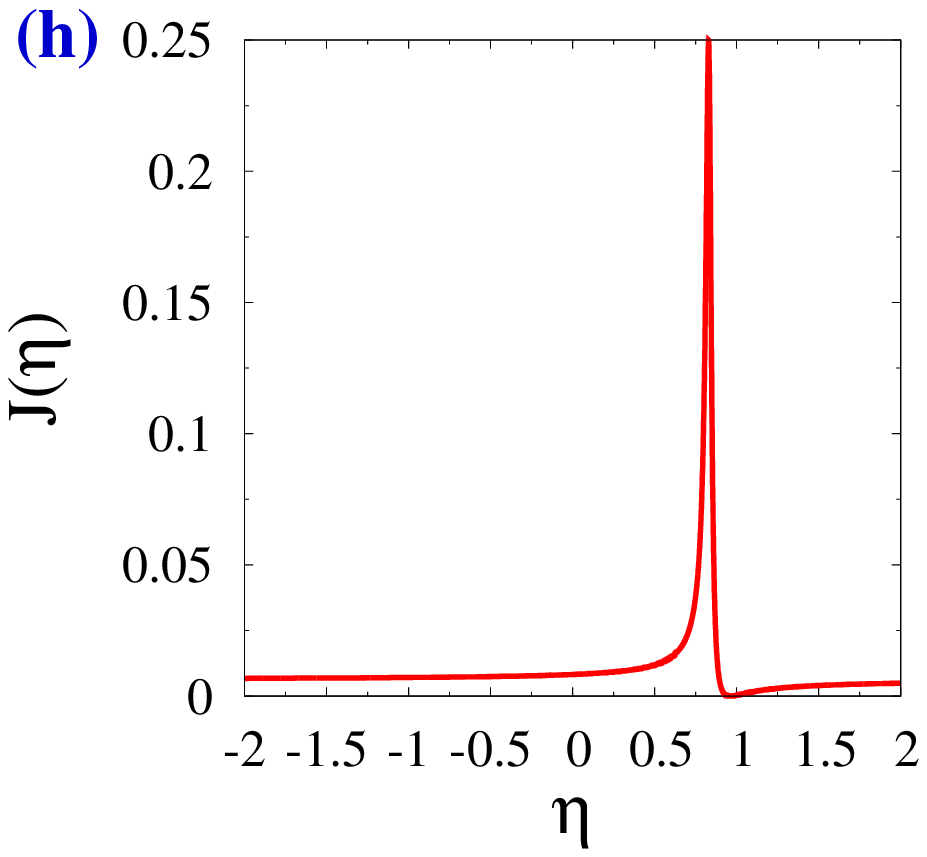}\includegraphics[height=3.5cm]{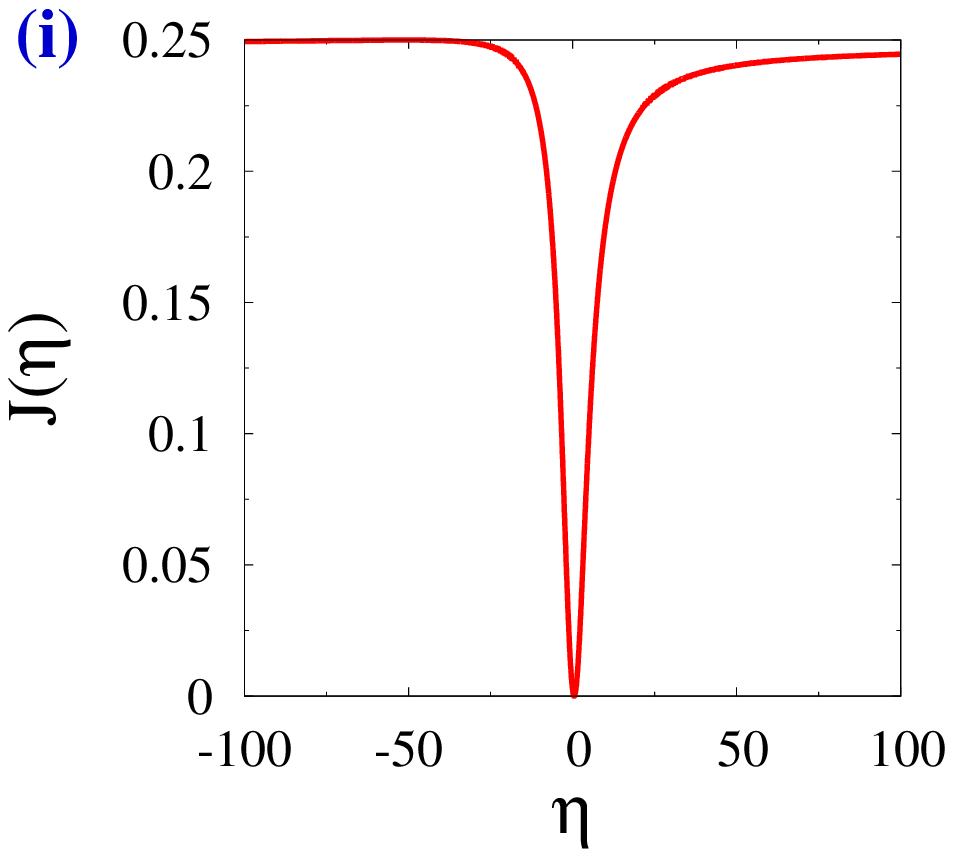}
    \caption{ Illustrations of the rich behavior of $J(\eta)$ at various limits. (a)-(b) Weak-coupling
      limit.  (a) The $r\to 0$ limit with $r=0.001$ and $q=0.5$. (b)
      The $q\to 0$ limit with finite $r$. We employ $q=0.001$ and
      $r=0.5$. (c)-(f) Several examples for achieving the tight-coupling limit $|q(1+r)|\to 2$ with
      maximum macroscopic efficiency. (c)
      $r=0.5$, $q=1.3332$ (d) $r=1$,  $q = 0.9999$, (e) $r=5$,
      $q=0.3332$, (f) $r=-1$, $q=10$. (g)-(i) Examples for the tight-coupling limit
      with maximum macroscopic output power. (g) $r=1$,  $q = 0.9999$, (h) $r=5$,
      $q=0.3332$, (i) $r=-1$, $q=10$. Note that the scales of the
      horizontal axes in these figures are different.}
    \label{fig:lim}
  \end{center}
\end{figure}

\subsection{IE. Width of efficiency distribution  under general conditions}

The width of efficiency distribution  at arbitrary $\alpha$ is found to be
\be
\sigma_\eta = \frac{2 \sqrt{2}
  |\alpha| (1 - q^2 r) (1 + \alpha^2 + \alpha q + \alpha q r)}{(1 + \alpha q)^2 \sqrt{
 4 - q^2 (1 + r)^2}} .
\ee
In Fig.~\ref{fig:wid-g} we display $\sigma_\eta$ as a function of the affinity parameter
$\alpha$ and the TRB parameter $r$ for $q=0.5$. We find that the width
of efficiency distribution  grows with $|\alpha|$; at $\alpha\sim 0$ it is very small.
Besides, it tends to very large values in approaching the tight-coupling limit.

\begin{figure}[]
\begin{center}
\includegraphics[height=4.2cm]{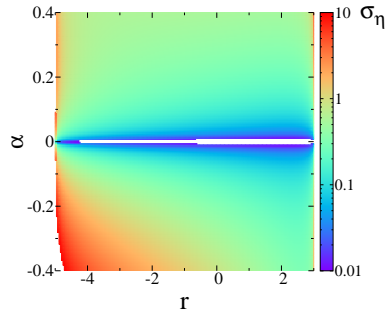}
\caption{ Width of efficiency distribution  $\sigma_\eta$ as a function of the affinity parameter $\alpha$ and the TRB parameter $r$ for $q=0.5$. The boundaries, $r=-5$ and $r=3$,
  satisfy the tight-coupling limit. White regions appear either
  because $\sigma_\eta$ is too small (around $\alpha=0$), or because it
  is too large (around $r=-5$ or $3$).}
\label{fig:wid-g}
\end{center}
\end{figure}

\subsection{IF. Width of efficiency distribution  under the maximum macroscopic
  power condition} 

Fig.~2(b) in the main text illustrates the behavior of
the width of efficiency distribution  $\sigma_\eta$ under the condition of maximum
macroscopic efficiency. Here we complement this result and
display in Fig.~\ref{fig:widthmp}(a) the behavior of $\sigma_\eta$ 
under the condition of maximum macroscopic power. We see in this figure
that the features of $\sigma_\eta$ are qualitatively
similar in both cases. In particular, at the small $q^2r$ limit,
$\sigma_\eta$ for both cases become quantitatively similar, which is
consistent with the understanding that the maximum macroscopic
efficiency and maximum macroscopic power conditions are close at this limit\cite{jiang}.
Nonetheless, we find that $\sigma_\eta$ diverges in the tight-coupling
limit for all $r$ except at $r=0, \pm 1$. In Fig.~\ref{fig:widthmp}(b)
we examine a particular case with $q=0.5$. The tight-coupling limit is
then reached for $r=-5$ or $r=3$, where $\sigma_\eta$ is shown to
diverge.

\begin{figure}[]
\begin{center}
\includegraphics[height=4.cm]{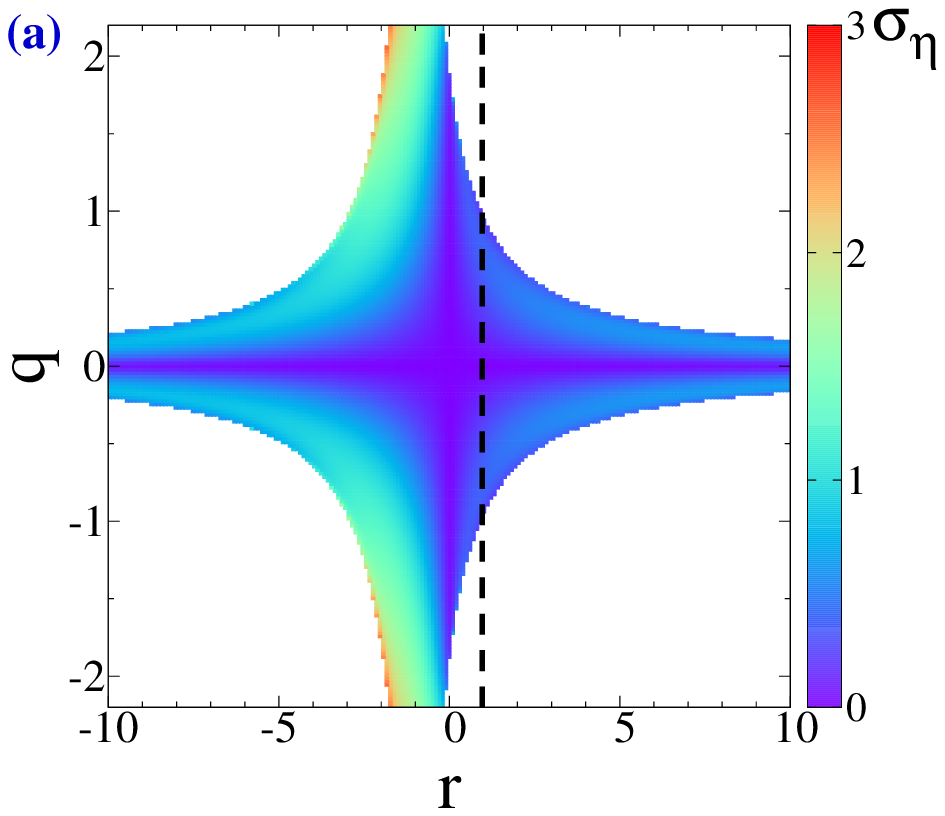}\includegraphics[height=4.cm]{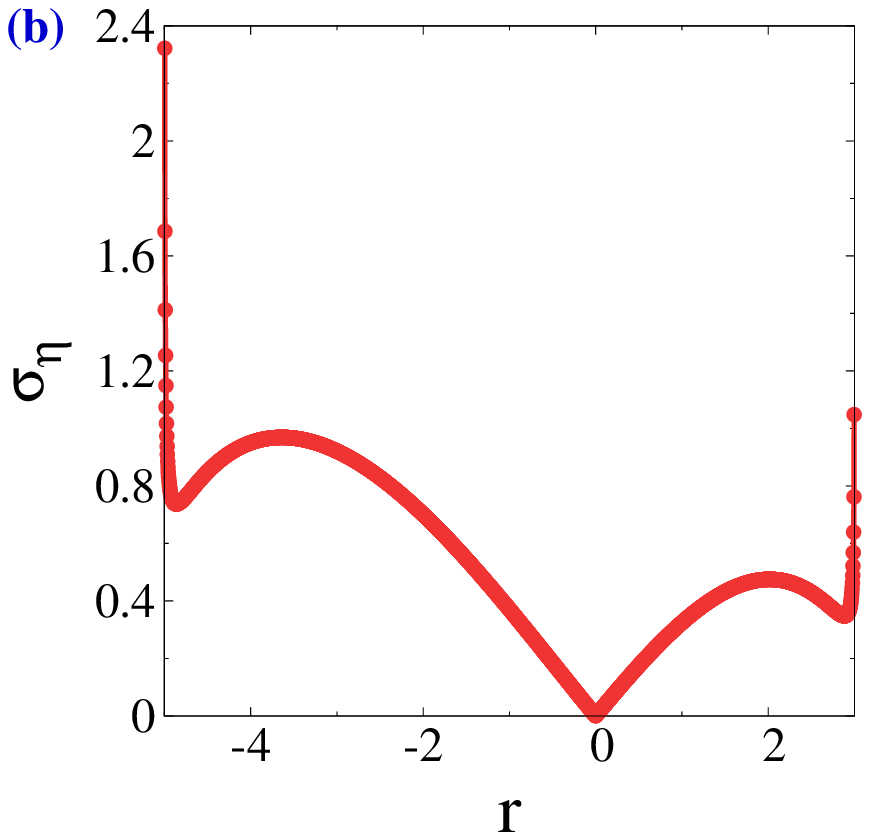}
\caption{ width of efficiency distribution  at the maximum macroscopic
  output power condition. (a) Width of efficiency distribution $\sigma_\eta$ as
  a function of $r$ and $q$. (b) Particular example of $\sigma_\eta$ with $r$ at $q=0.5$.}
\label{fig:widthmp}
\end{center}
\end{figure}

\section{II. Quantum mechanical bound on efficiency fluctuations}

We examine here the effect of the quantum mechanical bound, which was introduced in
Ref.~\cite{trb3}, on efficiency fluctuations.  Within our parameters
the bound comes up as $1 + q^2 r - q^2 - q^2 r^2 \ge 0$,
affecting the width of efficiency distribution 
fluctuation $\sigma_\eta$. The bound was obtained by considering a three-terminal  
Landauer-B\"uttiker model, and it is a direct result of the unitarity of the scattering matrix.
In Fig.~\ref{fig:QM} we plot (a) $\sigma_\eta$ under the maximum
macroscopic efficiency condition, (b) the least probable efficiency
$\eta^\star$, and (c) $\sigma_\eta$ under the maximum macroscopic power
condition. 
The quantum bound regularizes the divergent behavior of $\sigma_\eta$ under both
conditions, as well as the divergency of $\eta^\star$ for the maximum
power condition. This is because the above bound prohibits us from meeting the
tight-coupling limit, unless $r=1$, see $J(\eta)$ in Fig.~\ref{fig:lim}(d). 
However, note that the least probable efficiency can still diverge at the
following affinity parameter, even within the quantum mechanical bound,
\be
\alpha_c = \frac{q (1 -  r)}{2 - q^2 - q^2 r} .
\ee
At this value we receive positive average efficiencies and average
output power when $r>1$ or $r<-1$. Inspecting Fig.~\ref{fig:QM} further, we find that the
width of efficiency distribution  under both maximum macroscopic efficiency
and maximum macroscopic power conditions can become considerably large.
In addition, the least probable efficiency $\eta^\star$ can deviate significantly
from the Carnot efficiency. In fact, within
the maximum macroscopic efficiency condition, $\eta ^\star$ is allowed to diverge when
$r\to \infty$ \cite{trb2}. We note that the above bound goes to the
thermodynamic bound for $N$-terminal
Landauer-B\"uttiker conductors, when $N\to \infty$
\cite{trb3}. 
However, the quantum bound may not necessarily hold 
when genuine many body interactions and inelastic processes (e.g.,
electron-electron scattering and electron-phonon scattering) are taken into account.

\begin{figure}[]
  \begin{center}
    \includegraphics[height=4.2cm]{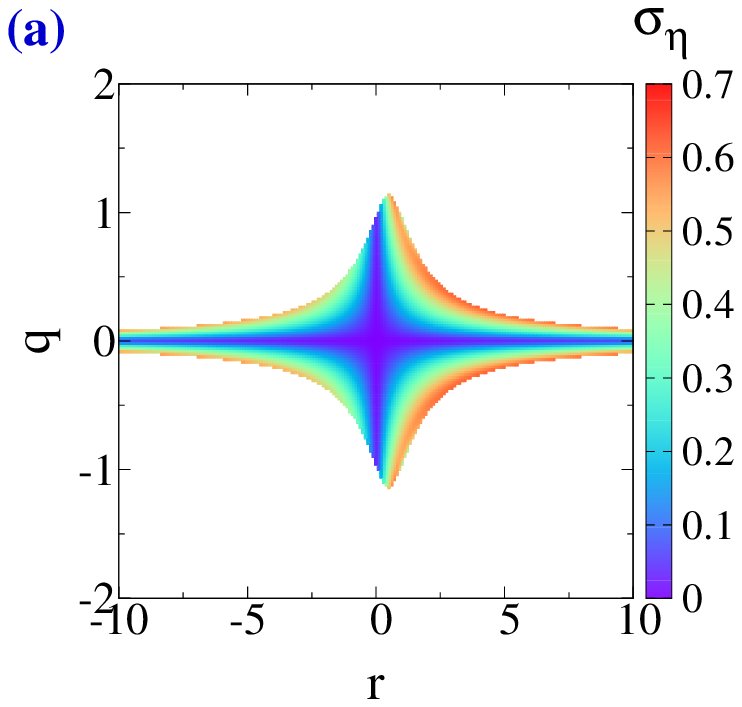}\includegraphics[height=4.2cm]{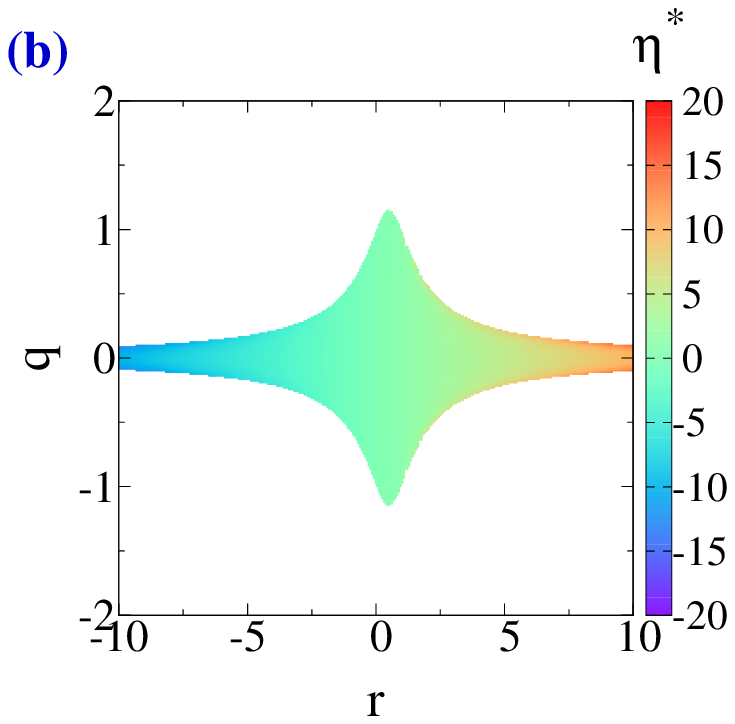}\includegraphics[height=4.2cm]{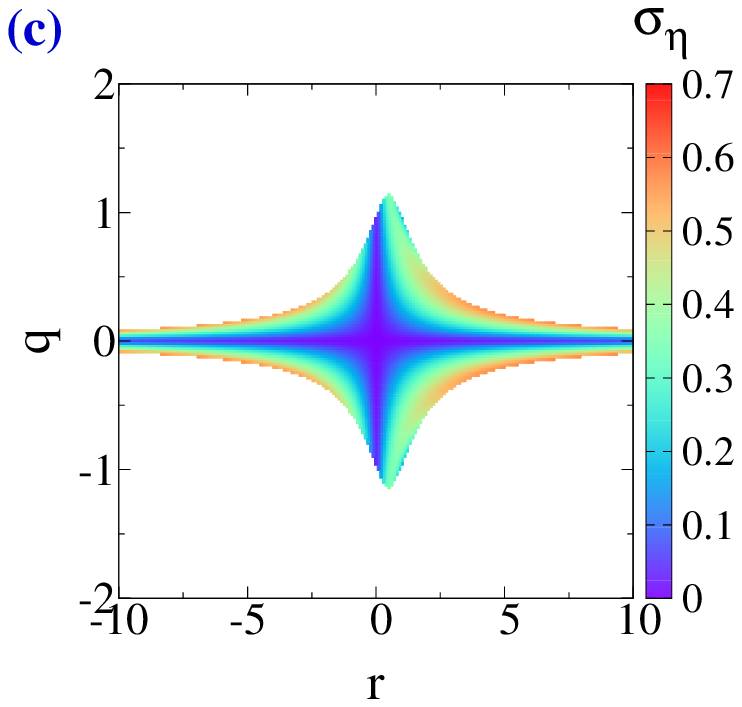}
    \caption{ Efficiency statistics while considering the quantum bound on transport
      coefficients. (a) Width of efficiency distribution 
      $\sigma_{\eta}$ under the maximum macroscopic efficiency
      condition. (b) Least probable efficiency $\eta^\star$ and (c)
      width of distribution $\sigma_\eta$ under the maximum
      macroscopic output power condition.}
    \label{fig:QM}
  \end{center}
\end{figure}

\section{III. Comparison between the Gaussian approximation and a Full
  counting statistics analysis for a triple-QDs thermoelectric model}

In this section we present a comparison between the LDFs as obtained
under the Gaussian approximation and a full counting statistics
analysis using a three-terminal thermoelectric model. A counting
statistics method including a probe terminal has been recently
developed by Utsumi {\sl et al.} (though the discussion there is focused on heat
currents fluctuations) \cite{utsumi}. In this approach the
LDF of efficiency is calculated through the cumulant-generating
function (CGF) defining counting fields for the particle number
$\xi_j$ and energy $\lambda_j$ in each reservoir $j=L,R,P$. The CGF of
the three-terminal system is given by \cite{utsumi}
\be
{\cal F}_{3t} = \int\frac{d\ome}{2\pi} \ln \det[\hat{1}
-\hat{f}(\ome)\hat{K}(\ome,\phi)] .
\ee 
Here 
\begin{align}
\hat{f} = {\rm diag}(f_L, f_R, f_P), \quad \hat{K} = \hat{1} -
e^{i\hat{\theta}} \hat{S}^\dagger(\ome, \phi) e^{-i\hat{\theta}}
\hat{S}(\ome,\phi), \quad \hat{\theta} = {\rm diag}(\ome \lambda_L +
\xi_L, \ome \lambda_R + \xi_R, \ome \lambda_P + \xi_P ) ,
\end{align}
with ``${\rm diag}$'' denoting a diagonal matrix, 
$f_j=[\exp(\frac{\ome-\mu_j}{T_j})+1]^{-1}$ ($j=L,R,P$) is the Fermi distribution
for $j-$th reservoir, and $\hat{S}$ is the S-matrix of the
triple-quantum-dot (QD) system. 

By integrating out the short-time
dynamics it was shown in Ref.~\cite{utsumi} that the effective CGF for particle and energy transport between
the L and R reservoirs (when the particle and energy currents
flowing out of the probe terminal are zero) is given by
\be
{\cal F}_{2t}(\lambda_L,\lambda_R,\xi_L,\xi_R, T_L, \mu_L, T_R, \mu_R) =
{\cal F}_{3t}(\lambda_L,\lambda_R,\lambda_P^\ast, \xi_L,\xi_R,
\xi_P^\ast, T_L, \mu_L, T_R, \mu_R, T_P^\ast, \mu_P^\ast) .
\ee
Here $\lambda_P^\ast$, $\xi_P^\ast$, $T_P^\ast$, and $\mu_P^\ast$ are
determined by the saddle-point equations
\be
\frac{\partial {\cal F}_{3t}}{\partial \lambda_P} = \frac{\partial
  {\cal F}_{3t}}{\partial \xi_P} = \frac{\partial {\cal F}_{3t}}{\partial T_P} =
\frac{\partial {\cal F}_{3t}}{\partial \mu_P} = 0 ,
\ee
which maximize the probability of processes with zero energy and particle
currents flowing out of the probe terminal.

In the linear-response regime, the effective two-terminal CGF can be
approximated by a second-order expansion in $\lambda_j$ and $\xi_j$ as
well as the affinities $(\mu_j-\mu)/T$ and $1/T- 1/T_j$ for $j=L,R$.
Due to particle and energy conservation, the counting fields for the
right reservoir can be regarded as redundant, hence we can set
$\lambda_R=\xi_R=0$. Furthermore we set $T_R=T$ and $\mu_R=\mu$.
The approximate two-terminal CGF is now given
by
\be
{\cal F}_{2t}(\vec{a}) = \frac{1}{2} \vec{a}^T\cdot \hat{{\cal R}}
\vec{a} ,
\ee
where 
\be
\vec{a}^T = (\lambda_L, \xi_L, A_1, A_2), \quad \hat{{\cal
    R}}_{\beta\beta^\prime} = \left . \frac{\partial^2 {\cal F}_{2t}}{\partial
  a_{\beta}\partial a_{\beta^\prime}} \right |_{\lambda_L=\xi_L=A_1=A_2=0} ,
\ee
with $A_1=(\mu_L-\mu_R)/T_R$ and $A_2=1/T_R-1/T_L$.

The matrix $\hat{{\cal R}}$ must be calculated from the second
derivative tensor of the three-terminal CGF at equilibrium with {\em
  vertex corrections} as shown in Ref.~\cite{utsumi},
\be
\left .\frac{\partial^2 {\cal F}_{2t}}{\partial a_{\beta}\partial a_{\beta^\prime}}\right |_{\lambda_L=\xi_L=A_1=A_2=0} =
\left . \left ( \frac{\partial^2 {\cal F}_{3t}}{\partial a_{\beta}\partial
    a_{\beta^\prime}}  -
\frac{\partial^2 {\cal F}_{3t}}{\partial a_{\beta}\partial b_{\gamma}}
U_{\gamma\gamma^\prime} \frac{\partial^2 {\cal F}_{3t}}{\partial
  b_{\gamma^\prime}\partial a_{\beta^\prime}} \right ) \right |_{\lambda_L=\xi_L=\lambda_P=\xi_P=A_1=A_2=A_3=A_4=0},\label{vert}
\ee
where $\vec{b} = (\lambda_P, \xi_P, A_3, A_4)$ with
$A_3=(\mu_P-\mu_R)/T_R$, $A_4=1/T_R-1/T_P$, and
$\hat{U}=\hat{B}^{-1}$ with
\be
B_{\gamma\gamma^\prime} = \left . \frac{\partial^2 {\cal
      F}_{3t}}{\partial b_{\gamma}\partial b_{\gamma^\prime}}\right
|_{\lambda_L=\xi_L=\lambda_P=\xi_P=A_1=A_2=A_3=A_4=0} .
\ee
We now change variables 
\be
i\lambda_L \to \eta \zeta (T_L-T_R)/T^2, \quad i\xi_L \to \zeta (\mu_L-\mu_R)/T ,
\ee
and obtain the LDF of efficiency fluctuations as\cite{me2}
\be
{\cal G}(\eta) = - \min_{\zeta} {\cal F}_{2t}(-i\eta \zeta, -i \zeta
(\mu_L-\mu_R)/T, A_1, A_2) 
\ee
for any given $A_1$ and $A_2$.

In Fig.~\ref{fig:cgf}(a) we plot the LDF $J(\eta)$ as obtained 
from the Gaussian approximation [Eq.~(\ref{eq:jeta-ap})] and that from the full counting
statistics method. The two functions perfectly match. In
Fig.~\ref{fig:cgf}(b) we plot the least probable efficiency
$\eta^\star$ calculated from the Gaussian approximation and the full
counting statistics method. Again, the results from the two methods
agree well with each other. These results
confirms the validity of our analysis in the main text based on the Gaussian approximation, a
consequence of the fluctuation theorem.

\begin{figure}[]
  \begin{center}
    \includegraphics[height=4.5cm]{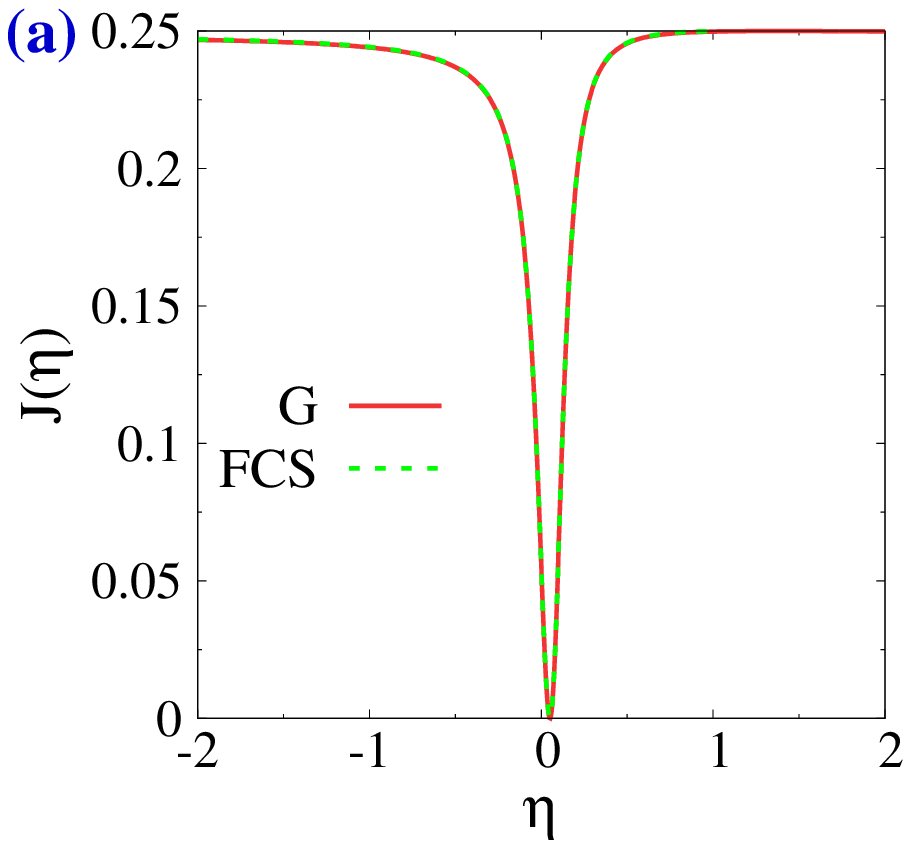}\includegraphics[height=4.5cm]{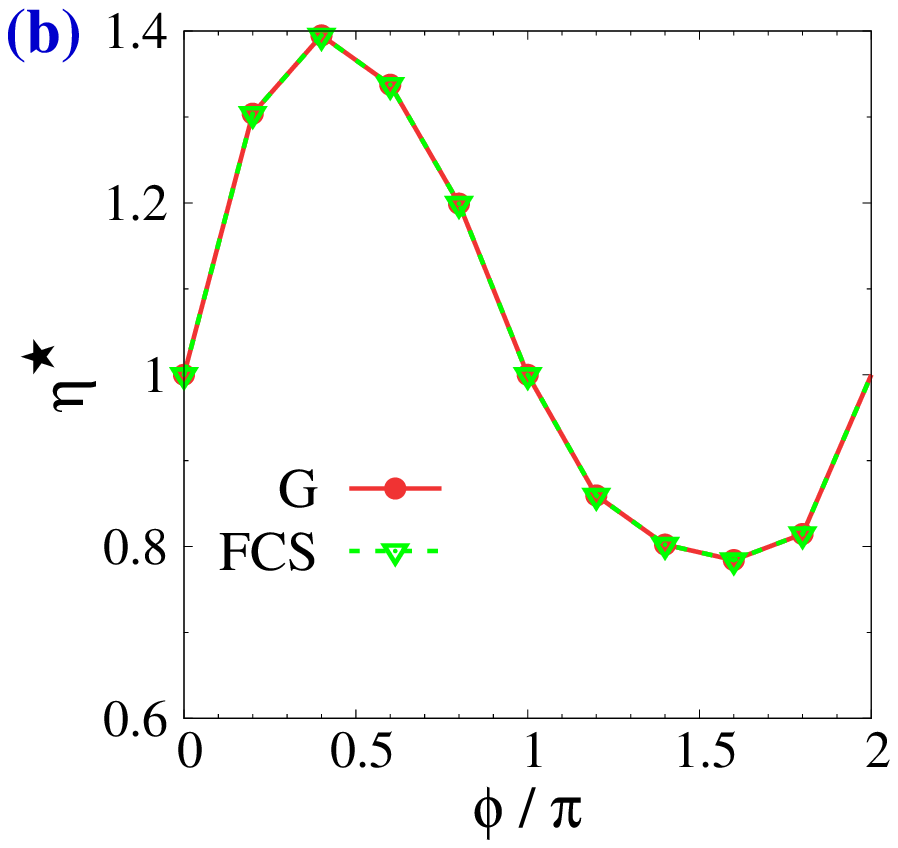}
    \caption{Comparison between $J(\eta)$ within the Gaussian approximation
      (labeled by ``G''), and from a full counting
      statistics analysis (labeled by ``FCS'') for the triple-QDs 
      thermoelectric model. (a) $J(\eta)$ for $\phi=\pi/2$. (b)
      $\eta^\star$ vesus $\phi$. Parameters are $E_1=1$, $E_2=0$,
      $E_3=1$, $\Gamma=1$, $t=0.4$, $T_L=1.125$, $T_R=1$,
      $\mu_L=- 0.01$, and $\mu_R=0$.}
    \label{fig:cgf}
  \end{center}
\end{figure}

\begin{figure}[]
 \begin{center}
   \includegraphics[height=4.cm]{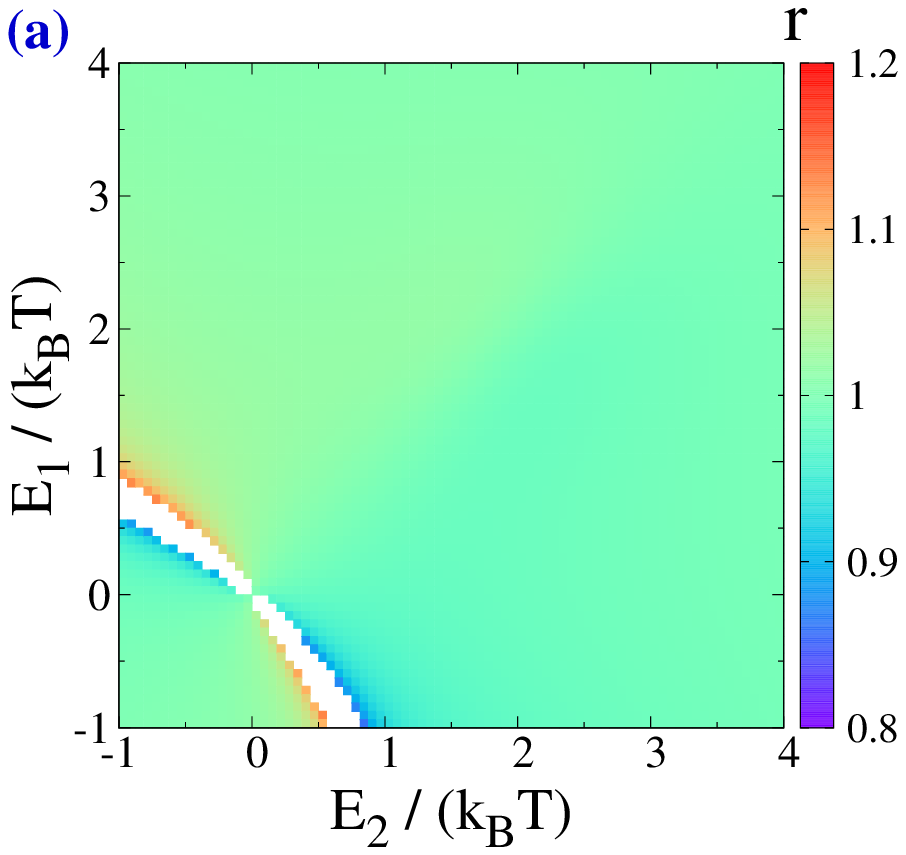}\includegraphics[height=4.cm]{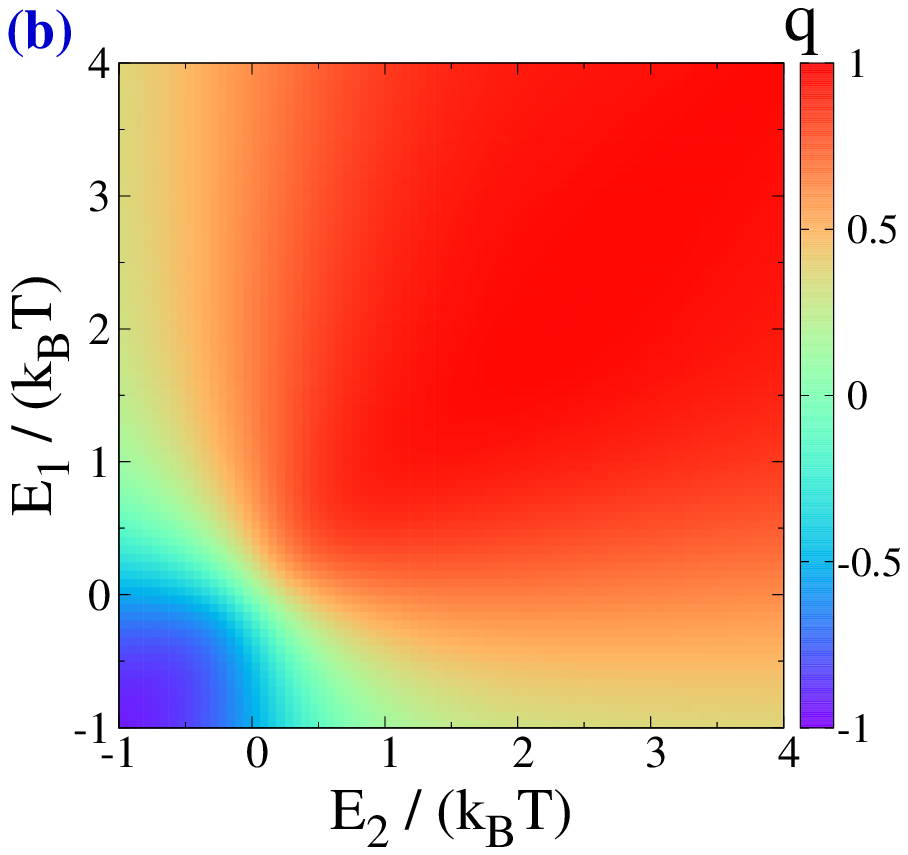}\includegraphics[height=4.cm]{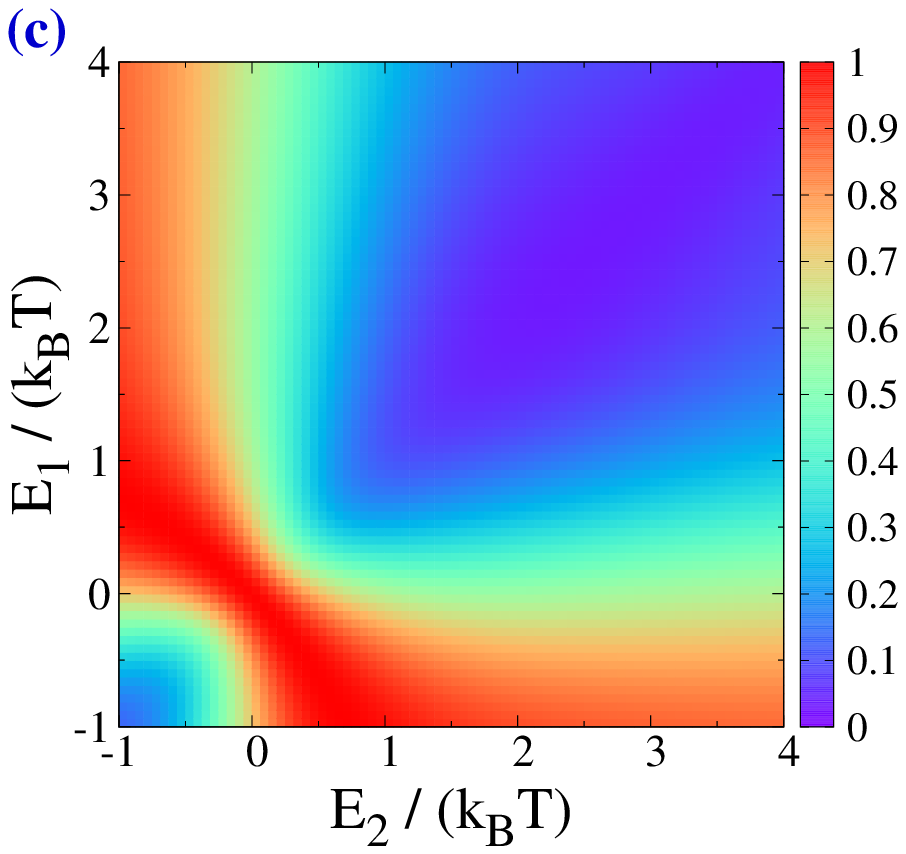}\includegraphics[height=4.cm]{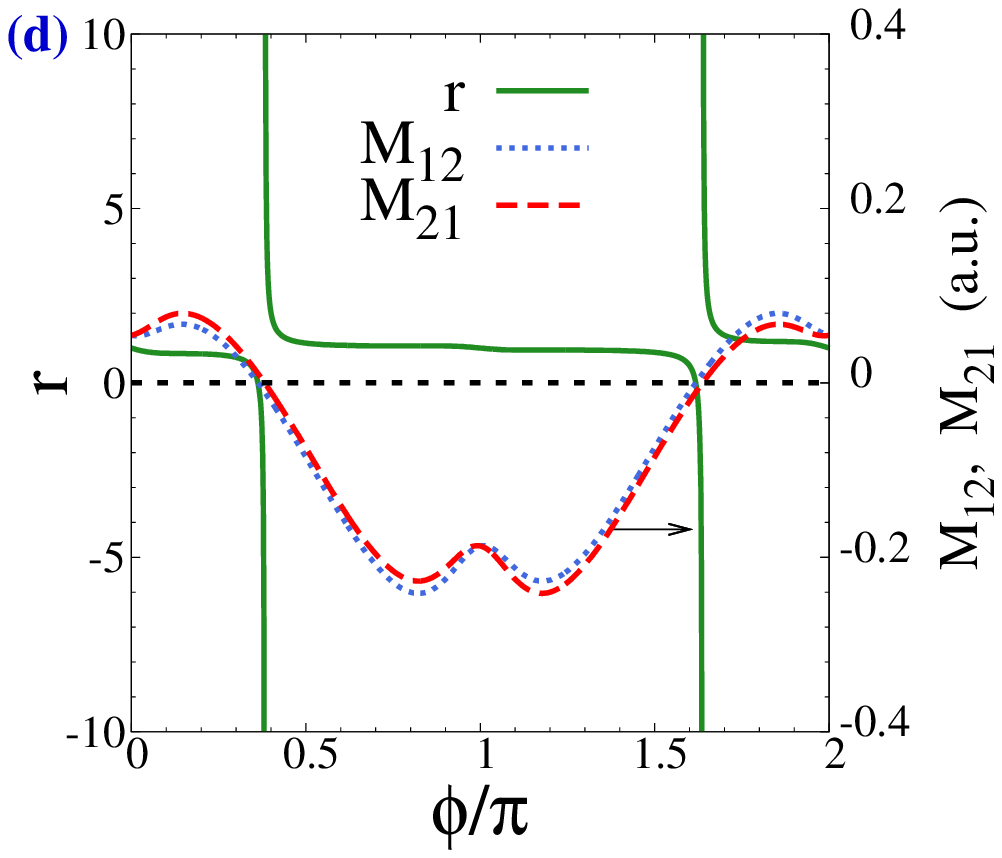}
   \caption{Time-reversal symmetry breaking in a 
     three-terminal triple QDs model for thermoelectrics. (a) TRB parameter $r$ and (b) degree
     of coupling $q$ as functions of the QDs energies $E_1$
     and $E_2$. (c) The function $1 + q^2 r - q^2 - q^2 r^2$ for the same values of energies as in panel (a), demonstrating that the quantum bound is satisfied. In (a)-(c) we used the following parameters: the QD connected to the probe terminal
     has energy $E_3=2$, $\phi=\pi/2$, $\Gamma=0.5$, and $t=-0.2$ 
     (energy unit is $k_BT=1$). (d) The example illustrates the
     appearance of $r\to \infty$ at a certain flux ($\phi$) value, when the parameters are chosen as
     $E_1=0.2$, $E_2=-0.2$, $E_3=1$, $\Gamma=0.5$, and $t=-1$.}
\label{fig:3qd-sup}
\end{center}
\end{figure}

\section{IV. Characterization of thermoelectric transport in a  triple-QDs  system}

We calculate linear transport coefficients for a three-terminal
thermoelectric model when the average thermal and electrical currents
flowing out of the probe terminal $P$ are zero. These conditions lead to
$\hat{M} = \hat{M}_{LL}^\prime - \hat{M}_{LP}^\prime 
\hat{M}_{PP}^{\prime -1} \hat{M}_{PL}^\prime$.
Here $\hat{M}^\prime$ is the transport matrix for the total three-terminal
system, i.e., $\vec{I}^\prime=\hat{M}^\prime\vec{A}^\prime$ where
$\vec{I}^\prime=(\vec{I}^\prime_L,\vec{I}^\prime_P)$ and
$\vec{A}^\prime=(\vec{A}^\prime_L,\vec{A}^\prime_P)$ with
$\vec{I}^\prime_\gamma=(I^\prime_{\gamma e}, I^\prime_{\gamma h})$
and $\vec{A}^\prime_\gamma=(A^\prime_{\gamma e}, A^\prime_{\gamma h})$
are  the currents and affinities for terminals $\gamma=L,P$. $I_e$ is
the charge current, $I_h$ is the heat current, and e.g., $A_{L
  e}=(\mu_L-\mu_R)/T_R$, $A_{L h}=1/T_R-1/T_L$. Linear
transport coefficients are calculated from the expression
\begin{equation}
\hat{M}^\prime_{\gamma\gamma^\prime} = \int \frac{d\ome}{2\pi}\, \left[\delta_{\gamma\gamma^\prime} -
  |S_{\gamma\gamma^\prime}(\ome,\phi)|^2\right] \left( \begin{array}{cccc}
    1 & \ome\\    
    \ome & \ome^2
  \end{array}\right) f_0(\ome) [1-f_0(\ome)],
\end{equation}
where $S_{\gamma\gamma^\prime}(\ome,\phi)$ ($\gamma,\gamma^\prime=L,P$) is the scattering matrix between
terminals $\gamma$ and $\gamma^\prime$, $\phi=2\pi\Phi/\Phi_0$. $\Phi$ and $\Phi_0$ are
the magnetic flux in our triple-quantum-dots (QDs) system and the flux
quantum, respectively. The Fermi distribution $f_0(\ome)=[\exp(\frac{\ome}{T})+1]^{-1}$
corresponds to an equilibrium state with the chemical
potential set at $\mu=0$. Onsager reciprocal symmetry
originates from the symmetry $S_{\gamma\gamma^\prime}(\ome,\phi)=S_{\gamma^\prime\gamma}(\ome,-\phi)$.

The transmission function is obtained from the relation 
$\hat{S}(\ome,\phi)=-\hat{1}+i\Gamma \hat{G}^r(\ome)$. Here $\hat{1}$
is a $3\times 3$ identity matrix and $\hat{G}^r(\ome)=[(\ome +
i\Gamma) \hat{1} - \hat{H}_{qd}]^{-1}$ is the retarded Green's
function of the QDs. The hybridization energy 
$\Gamma=2\pi\sum_{k}|V_{k}|^2\delta(\ome-\vep_{k})$ is assumed  
to be a constant (independent of energy) for all three QDs.

We calculate the transport coefficients and then determine the TRB
parameter $r$ and the degree of coupling $q$. In
Fig.~\ref{fig:3qd-sup} we plot these parameters against $E_1$ and $E_2$. We further confirm the
``quantum bound'' on linear transport coefficients discovered by
Brandner {\sl et al.} for three-terminal TRB conductors\cite{seifert},
which using our parametrization casts into the form as $1 + q^2 r - q^2 - q^2 r^2 \ge 0$.
In Fig.~\ref{fig:3qd-sup}(c) we plot it and show that it is always greater than zero in our parameter region.
To illustrate the divergence of $r$, we plot it in Fig.~\ref{fig:3qd-sup}(d)
as a functions of the magnetic flux $\phi$. We find that $r$ diverges when $M_{21}$ goes to zero
with a finite $M_{12}$. This happen when $\phi\simeq 0.4\pi$ or $\phi \simeq 1.6\pi$.

\end{widetext}

\end{document}